\documentclass[fleqn,10pt]{wlscirep}
\usepackage[utf8]{inputenc}
\usepackage[T1]{fontenc}

\title{Metastability, atmospheric midlatitude circulation regimes and large-scale teleconnection: a data-driven approach}

\author[1,*]{Dmitry Mukhin}
\author[1]{Roman Samoilov}
\author[2]{Abdel Hannachi}
\affil[1]{Institute of Applied Physics of RAS, Nizhny Novgorod, 603950, Russia}
\affil[2]{Stockholm University, Department of Meteorology, Stockholm, SE-106 91, Sweden}

\affil[*]{mukhin@ipfran.ru}

\begin{abstract}
The low-frequency variability of the mid-latitude atmosphere involves complex nonlinear and chaotic dynamical processes posing predictability challenges. It is characterized    by sporadically recurring, often long-lived patterns of atmospheric circulation of hemispheric scale known as weather regimes. The evolution of these circulation regimes   in addition to their link to large-scale teleconnections can help extend the limits of atmospheric predictability. They also play a key role in sub- and inter-seasonal  weather forecasting. Their identification and modeling remains an issue, however, due to their intricacy, including a clear conceptual picture. In recent years, the concept of metastability has been developed to explain regimes formation. This suggests an interpretation of circulation regimes as communities of states in which the atmospheric system remains in their neighborhood for abnormally longer than typical baroclinic timescales. Here we develop a new and effective method to identify such communities by constructing and analyzing an operator of the system's evolution via hidden Markov model (HMM). The method makes use of graph theory and is based on probabilistic approach to partition the HMM transition matrix into weakly interacting blocks -- communities of hidden states -- associated with regimes. The approach involves nonlinear kernel principal component mapping to consistently embed the system state space for HMM building.  Application to northern  winter hemisphere using geopotential heights from reanalysis yields four persistent and recurrent circulation regimes. Statistical and dynamical characteristics of these circulation regimes and surface impacts are discussed. In particular, unexpected high correlations are obtained with EL-Nino Southern Oscillation and Pacific decadal oscillation with lead times of up to one year. 

\end{abstract}
\begin{document}

\flushbottom
\maketitle

\thispagestyle{empty}

\section{Introduction}

Humans have, since the dawn of civilization from the Babylonian to present day, sought to  predict the weather far in the future. The theory of modern mathematical models of nonlinear dynamical systems show that weather cannot be forecast accurately beyond $1-2$ weeks ahead. However, this is not the full story. In fact, the atmosphere exhibits complex variation on a wide spectrum  of spatio-temporal scales ranging from synoptic to decadal and longer time scales originating from internal nonlinear  dynamics and also from the boundary conditions such as sea surface temperature (SST) variation.  These sources can contribute to extending the chaotic predictability limits to include subseasonal-to-seasonal (S2S) timescales.

Low-frequency variability (LFV) of the atmosphere, spanning spatio-temporal scales  greater   than typical synoptic and baroclinic timescales is poorly understood and is difficult to  predict. Extratropical LFV, in particular, involves recurring  and persistent large-, e.g. hemispheric-scale structures. These structures of atmospheric circulation are often referred  to as teleconnection patterns or regimes and are known to have profound impact on weather predictability on intraseasonal and longer timescales \cite{hannachi2017,ghil2019}. With such scales, the ensuing circulation anomalies strongly modulate the tracks and intensity of synoptic disturbances in the atmosphere, operating like waveguides, and hence determine the long-term behavior of weather over large areas. The modeling and analysis of the mid-latitude climate is closely related to recurring circulation regimes. Both the identification and the understanding of their dynamics continue to be a debatable issue due to the complexity and nonlinearity of the   problem beside the unsettled concept of regimes.
%and many different methods used. 

Quasi-geostrophic theory of large-scale atmospheric circulation shows that LFV is essentially the result of nonlinear dynamics \cite{Marshall1992}.  Experimental analyses based on quasi-geostrophic barotropic and baroclinic models reveal the high-dimensional nature and chaotic dynamics of the LFV atmospheric flow \cite{Vannitsem1997}. There   is suggestion in the literature \cite{Gritsun2017} that in an infinite set of saddle-invariant manifolds of a chaotic attractor, the most stable of those manifolds form  the skeleton of the dynamics, and sporadic adherence of the system state in their neighborhood can cause regime persistence. This nonlinear perspective leads to the understanding of regimes as metastable regions in the system's phase space where the system slows down. To consistently adopt this paradigm of {\it persistent} regimes,   an extra requirement on regime detection methodology has to be added, namely, it  should account for temporal relationships of observed states rather than just their similarity in a selected projection of phase space. A number of existing methods \cite{Hannachi21} for identifying recurrent patterns, e.g. based on PDF decomposition or partitioning networks of similarity, do not satisfy this requirement. 

An appropriate and elegant way to address this issue is to construct and analyze an evolution operator of the system using hidden Markov model (HMM) \cite{Murphy2012, Franzke2008}. A Conventional Markov chain is a stochastic model based on 'observed' states of the chain. A HMM \cite{Murphy2012, rabiner1989} is an extension of the Markov chain by including other ({\it hidden}) states, which generate the observed sequence. Because of the finite and short-memory representation of the system, HMM is  a consistent method for detecting metastability as well as identifying the basins of metastable states. Essentially,  with HMM nonlinear effects, like metastability and trace of multiple equilibria, can be captured by linear transfer operator pretty much like dynamic mode decomposition (DMD) \cite{Brunton2022}. HMM has been used to study atmospheric regimes via spectral analysis of the Markovian operator \cite{Franzke2008,Springer2024}. 

Inspired by ideas from \cite{Franzke2008}, we elaborate a new methodology for constructing and analyzing HMM aimed at detecting persistent and recurrent circulation regimes determined as metastable communities of hidden states. We derive and use a method of partitioning the HMM transition probability matrix into blocks, providing significantly low probabilities of transitions between them.  The method is based on maximizing a cost-function for a given partition and has a clear probabilistic meaning. In fact, the approach finds that partition with abnormally high mean probability that the state will remain within its community, i.e. persistent. We show that this optimization problem effectively reduces to the Newman iterative algorithm, widely used in graph theory for graph decomposition \cite{Newman2006}. We pay special attention to the significance test, which rejects the null hypothesis that the detected communities could have been obtained by chance from a sample of a randomized process with similar spectral properties.

Another key part of the methodology is the choice of an embedding space, i.e. a set of dynamic variables for dimension reduction and model construction. This is of paramount importance because it determines the projection of the phase space in which the model operates. The dimension of this space is desired to be as low as possible to ensure  stable results during the HMM learning process, but at the same time it should be informative enough for regime detection and identification. Kernel principal component analysis (KPCA) \cite{Scholkoph1998,Hannachi21} is considered here, which belongs to the class of nonlinear data transformation, and has been shown to enable a reliable and robust separation of states based on similarity measure in a space of just few leading components. Recent works \cite{Mukhin22,HannachiIqbal2019a}  used it to study  recurrent patterns of the mid-latitude atmospheric circulation. In this manuscript, KPCA serves as  a prepossessing step for the HMM to get both recurrent \cite{hannachi2017} and persistent circulation flows. 

The paper is organized as follows. In Section ~\ref{sec:methods} we describe the methodology of  HMM building, regimes  detection and analysis, with an illustration using a low-order chaotic dynamical system. Section ~\ref{sec:results}  presents the properties of circulation regimes obtained   from historical climate data. Finally, in Section ~\ref{sec:discussion}, we summarize and discuss advantages and prospects of the presented methodology and results. Model examples together with additional details on the methodology as well  as additional illustrations are detailed in the Supplementary Material.

\section{Methods}
\label{sec:methods}
\subsection{Hidden Markov model}

In the present setup the observed atmospheric variability is represented by time series of some dynamic variables $\mathbf X=(x_1,x_2,\dots,x_d)$. 
For example, in case of spatially distributed data, as is the case here, the embedding space for $\mathbf X$ can be constructed by means of projecting the data onto a lower dimensional space such as the linear empirical orthogonal functions (EOFs) or using a more sophisticated nonlinear functions, e.g., kernel principal components (see Sec.~\ref{kpca}). 
We assume that the evolution of these variables is driven by transitions between a finite number of hidden states of the system, $J=1,2,\dots,K$, where $J$ denotes a state number. 
The transitions are random and are modeled by a discrete-time Markov chain defined by the transition probability matrix $\mathbf Q_{ij}=p(J_i|J_j)$, which is a stochastic matrix, i.e. has a unit sum along each column ($\sum\limits_{i}{\mathbf Q_{ij}}=1$).  
Each hidden state is connected with the dynamic variables via emission probabilities $P(\mathbf X|J)$. 
Here we use Gaussian probability density function (PDF) for the emissions:
\begin{equation}
\label{emissions}
P(\mathbf X|J)=\frac{1}{\sqrt{(2\pi)^d |\Sigma_J|}}
\exp\left(-\frac{1}{2}(\mathbf X - \mathbf a_J)^T\Sigma_J^{-1}(\mathbf X - \mathbf a_J)\right),
\end{equation} 
although more complex PDFs can be employed, e.g., mixture models. 

The described HMM is fully determined by its parameters $\mu=(\mathbf Q, \Sigma, \mathbf a)$, which are to be estimated from the observed time series $(\mathbf X_1, \mathbf X_2,\dots,\mathbf X_N)$. The likelihood function for HMM contains the sum over huge number of possible sequences of the hidden states                                 $\mathbf S_1, \mathbf S_2, \ldots$, where each $\mathbf S_k$ contains $N_k$ hidden states                          $\mathbf S_k=\{J_1,J_2,\dots,J_{N_k}\}$, with $J_{N_k}<N$:
\begin{equation}
\label{lkh}
P(\mathbf X_1,\dots,\mathbf X_N |\mu)=\sum\limits_k {P(\mathbf X_1,\dots,\mathbf X_N, | \mathbf S_k,\Sigma,\mathbf a)}P(\mathbf S_k|\mathbf Q),
\end{equation}                
and hence, cannot be calculated and maximized directly. Here we use Baum-Welch algorithm \cite{Baum1970} that provides convergence to a local maximum of Eq. (\ref{lkh}). This is an iterative EM-algorithm, based on repeating two steps: (i) estimating the state occupation probabilities (E-step) and (ii) estimating the HMM parameters $\mu$ using the occupation probabilities (M-step). Thus, at the end of this procedure, we obtain the estimates of transition probabilities, position and orientation of emission functions in the phase space of dynamic variables $\mathbf X$, in addition to the occupation probability of the observation $\mathbf X_t$ in the hidden state $J=j$: $P(J=j|\mathbf X_t)$.

Since we are to train the HMM from natural stochastic time series, we will almost surely obtain a Markov chain that does not have both fully closed communities (classes) of states and pure periodic states. This guarantees the existence of a unique stationary distribution $\pi$ to which the Markov process converges starting from any initial distribution ${\mathbf q}$:
\begin{equation}
\label{stat}
\lim\limits_{k\to\infty}{\mathbf Q^k} {\mathbf q}=\pi,
\end{equation}   
or, equivalently, $\pi$ is the unique right eigenvector of the matrix $\mathbf Q$ corresponding to its largest eigenvalue which is exactly equal to $1$: $Q\pi=\pi$.  One more assumption we use in Eq. (\ref{stat}) is that   the transition matrix does not depend on time, i.e. a stationary process. Although this assumption seems relevant in case of quite short atmospheric time series (after removing the regression on the CO2 trend; see the result section), it should be relaxed when studying a nonlinear response of the atmosphere to external forcing. 

The number of hidden states $K$ is a structural parameter of the HMM, that determines the model resolution. Generally speaking, it is recommended to set this parameter as large as possible, provided that it is adequate to available statistics, i.e. each hidden state is reasonably well visited in the observed sample. While a too-coarse model with unreasonably small $K$ may miss important features of the metastable state space, an overfitted model with too large $K$ will have a nearly binary transition matrix $\mathbf Q$ that describes the unique given sample rather than the underlying system, i.e. overfitting. Since the Baum-Welch algorithm used for learning HMM allows us to calculate the likelihood Eq. (\ref{lkh}), the optimal $K$ can be roughly estimated based on the conventional Akaike information criterion (AIC). The idea behind is that the large-scale structure of the metastable regimes we are trying to find should be insensitive to $K$ when $K$ greatly exceeds the number of regimes (e.g., saturation), and any value near the AIC optimum provides a good model for regime detection. 
%Please see Supplementary Material for further details.     
    
\subsection{Modularity, graphs and persistent regimes}
\label{sec:pers_reg}

According to Franzke et al.\cite{Franzke2008}, an atmosphere circulation regime is associated with a metastable state of the system, i.e. a region of slow evolution of the trajectory in the system's phase space.   
With the HMM representation of the atmospheric circulation, a regime can be sought as a block (a subset of the hidden states) of the transition matrix with small probabilities of transitions to the states out of the block. 
Decomposing the full transition matrix $\mathbf Q$ into such blocks, some sort of tiling, gives us a reduced matrix of transition probabilities between the basins of stability in the space of hidden states. 
Going this way, we treat the HMM hidden states as microstates of the system \cite{Springer2024}, meanwhile a regime is determined as a community of the microstates. 
Such an approach helps reduce the influence of a specific sample on the result. 
The coverage of the data space by the hidden states, their size and location, are highly sensitive to the data sample we have. For example, the longer the time series on which the model is trained, the more hidden states we can expect provided that each of them is well visited by the observations. In other words, the microstates themselves reflect particular observations rather than the  dynamical properties of the system and lack physical interpretation. In our study, they serve only as a support for representing fundamental, data-independent invariant structures in the system's phase space, that underlie the metastable regimes.              

Now we can safely assume that a sign of metastability in a Markov chain appears when there exists a partition       $\{A_k\}$ of the hidden states such that the average probability of remaining in a given subset $A_k$ is larger, compared to that in a completely randomized chain having the same stationary distribution $\pi$. Thus, to find the best partition $\{A_k\}$, we should maximize the quantity:
\begin{eqnarray}
\label{measure}
\left<P_Q(J\in A_k|J\in A_k)-P_{Q^\infty}(J\in A_k|J\in A_k)\right>_{A_k} \nonumber \\
=\sum\limits_{k}{\left[P_Q(J\in A_k|J\in A_k)-\pi(A_k)\right]\pi(A_k)},
\end{eqnarray}
where $P_Q(a|b)$ is the probability of an event $a$ given an event $b$ at the previous state in the Markov chain $Q$; $\pi(A_k)$ is the stationary probability of a subset $A_k$, i.e. $\pi(A_k)=\sum\limits_{i\in A_k}{\pi_i}$; the chain $Q^\infty$ is the white-noise process with the stationary distribution $\pi$, which is generated by the transition matrix $\mathbf Q^\infty$ having identical columns given by $\pi$. The obtained largest value of Eq. (\ref{measure}) should be positive to consider the partition viable, otherwise we conclude that there are no persistent regimes states.  

Noting that 
\begin{equation}
P_Q(J\in A_k|J\in A_k)=\frac{1}{\pi(A_k)}\sum\limits_{i,j\in A_k}{Q_{ij}\pi_j}, \nonumber
\end{equation}         
we can rewrite Eq. \ref{measure} in the following form:
\begin{equation}
\label{modularity}
M=
\sum\limits_{A_k}{
\sum\limits_{i,j\in A_k}{\left[Q_{ij}\pi_j-\pi_i\pi_j\right]}}
=\sum\limits_{i,j}{\left[Q_{ij}\pi_j-\pi_i\pi_j\right]g_{ij}}
=\sum\limits_{i,j}{B_{ij}g_{ij}},
\end{equation}
where $g_{ij}=1$ if hidden states $J=i$ and $J=j$ belong to the same subset $A_k$ and $g_{ij}=0$ otherwise. 
Written this way, and this is the novelty here, the expression $M$ in Eq. (\ref{modularity}) is precisely similar to the modularity introduced by
Newman \cite{Newman2006}, which stems from graph theory \cite{Chung1997, lafon2006}. The modularity was previously used by \cite{Mukhin22} for detecting atmospheric regimes in connection with recurrence networks. One of the attractive feature of this formulation, Eq. (\ref{modularity}), is the possibility of using the iterative splitting algorithm to maximize $M$ simply via solving for an eigenvalue problem. The algorithm proceeds by iteratively 
splitting each subset $A_k$ into two communities so that each split provides the maximal positive increment of the modularity $\Delta M$, until indivisible subsets are
obtained.  Splitting of some subset $A$ into two groups can be represented by a classifier -- a vector $\mathbf s$ of dimension $|A|$ (size of $A$) whose elements $s_i =-1$ (for the first group) and $s_i = 1$ (for the second group), $i=1, \ldots |A|$). 
Now denoting by $\mathbf B^{(A)}$ 
the submatrix of $\mathbf B=(B_{ij})$ obtained by selecting the elements of $\mathbf B$ with the indices $i,j\in A$,
it can be shown that the increment $\Delta M$ of the whole modularity, after this split, takes the form:
\begin{eqnarray}
\label{split}
\Delta M=\frac{1}{2}\sum\limits_{i,j}{(B_A)_{ij}(s_is_j+1)}
=\frac{1}{2}\mathbf s^T\mathbf B_A\mathbf s, \\
\nonumber
(B_A)_{ij}=B^{(A)}_{ij}-\delta_{ij}
\sum\limits_{k}{B^{(A)}_{ik}},
\end{eqnarray} 
where $\delta_{ij}$ is the Kronecker delta. The algorithm is initiated by setting $A$ equal to the whole set of hidden states, that is $\mathbf B_A=\mathbf B$.

It is worth noting that the matrix $\mathbf B_A$ is symmetric only when the Markov chain obeys the detailed balance condition, i.e. $Q_{ij}\pi_j=Q_{ji}\pi_i$. However, this is not true for a general dynamical system and is not required in the HMM training procedure. The interesting point here, another advantage of this formulation, is that in the general non-symmetric situation, finding the binary vector $\mathbf s$ that maximizes $\Delta M$, makes use of the fact that the quadratic form in Eq. (\ref{split}) depends only on the symmetric part of the matrix $\mathbf B_A$, that is $\mathbf H=\frac{1}{2}\left(\mathbf B_A+\mathbf B_A^T\right)$, and the change in the modularity can simply be expressed in terms of the eigen-elements of $\mathbf H$:
\begin{equation}
\label{contrib}
\Delta M_A=\frac{1}{2}\mathbf s^T\mathbf H \mathbf s=\frac{1}{2}\lambda_1(\mathbf w^T_1\mathbf s)^2+\dots,
\end{equation} 
where $\lambda_1$ is the largest eigenvalue of $\mathbf H$ and $\mathbf w_1$  the corresponding eigenvector.  Eq. (\ref{contrib}) tells us that the binary vector maximizing the modularity increment $\Delta M_A$ is precisely given by the sign of $\mathbf w_1$, $\mathbf s=sign( \mathbf w_1)$.
In other words, the optimum vector $\mathbf s$ is the one whose $i'th$ component $s_i$ is 1 if the corresponding component of $\mathbf w_1$ is positive and -1 otherwise. 
A key conclusion follows from this, namely, if the matrix $\mathbf H$ has positive eigenvalues,  two new subsets defined by the vector $\mathbf s$ emerge, provided
that $\Delta M_A>0$. Otherwise, its largest eigenvalue is zero, since this matrix has the zero sum over each row or column, and no further splitting of the community $A$ is possible. 

As with most HMM, the probability of staying in each regime $A_k$ is relatively high, but the residence time     of the system in individual hidden states belonging to $A_k$ can differ. Eq. (\ref{contrib}) shows that the absolute values $|{w_1}_i|$ of the $i'th$ component of $\mathbf w_1$ determines the contribution of states with the corresponding classifiers $s_i$ to the modularity increment. This means that the system more often falls into states corresponding to large          $|{w_1}_i|$, and hence, this value can be treated as a measure of state stability. The most stable states form the cores of the obtained regimes and constitute the skeleton of metastable structure of the system phase space. The metastable states of dynamical systems determine in fact the structure of the system's attractor. 

The splitting process continues until the increment $\Delta M_A$ ceases to be positive for the
subset $A$. In practical terms a threshold is used for $\Delta M_A$ in order to eliminate insignificant splits. The significance test is 
based on surrogates obtained by random block shuffling of a time series generated by the HMM preserving main spectral properties of the original process (see Supplementary Material). 
Given the fixed emission PDFs, Eq. (\ref{emissions}), estimated from data, an ensemble    of transition matrices is obtained by training the HMM on the ensemble of surrogates.  
These surrogates preserve the stationary distribution $\pi$, but may not preserve
autocorrelations at large lags.   
The ensemble of the modularity increments $\Delta M_A$ based on the surrogate matrices is then  used to construct a confidence interval of $\Delta M_A$ from the original sample. 
Specifically, when a split is carried out of some community with the original $\mathbf Q$, the  same block in every surrogate matrix is also split to obtain the null distribution of increments. Thus, the value of increment on the right tail of the surrogate ensemble, that corresponds to the significance level, is used as a threshold for accepting $\Delta M_A$.            

\subsection{Time step selection}
\label{sec:timestep}

An important aspect of HMM is the choice of time step. For example, using daily data successive states are very likely to be highly correlated. As a result, the diagonal elements of the transition matrix will be dominant, reflecting high persistence  in the hidden states, and making partition difficult and the results prone to noise. Traditionally, this weakness is addressed by  adjusting the time step via smoothing and/or resampling the data.  
But the HMM provides another solution to analyze the Markov operator across $L$ days, that is,   to decompose the transition matrix $\mathbf Q^L$ (instead of the one-step $\mathbf Q$) where $L$ represents the spatiotemporal resolution of regime determination. This approach is akin to  diffusion maps, see e.g. \cite{lafon2006, nadler2006, belkin2003}, which selects the step size of the Markov process to determine the scale of the underlying structure.
Moderately large $L$ reduces regime sensitivity to noise. However, quite large $L$ can lead to matrix degeneracy, eventually ending up with rank-1 transition matrix with columns composed of the stationary distribution $\pi$.
We can expect that persistent states are reliably detected when $L$ is comparable to the characteristic timescale of the regimes, hence the need to look for a range of $L$ for which the regimes do not change significantly. 
It is worth noting that by adjusting the time step via $L$, we do not distort the original data as with smoothing and/or resampling, but rather involve all daily observations in model learning. A schematic representation of the method is sketched in Fig. ~\ref{lor_scheme}a.

\subsection{Illustration with a low-order chaotic model}
\begin{figure}[ht]
	\centering 
	\includegraphics[width=0.8\textwidth]{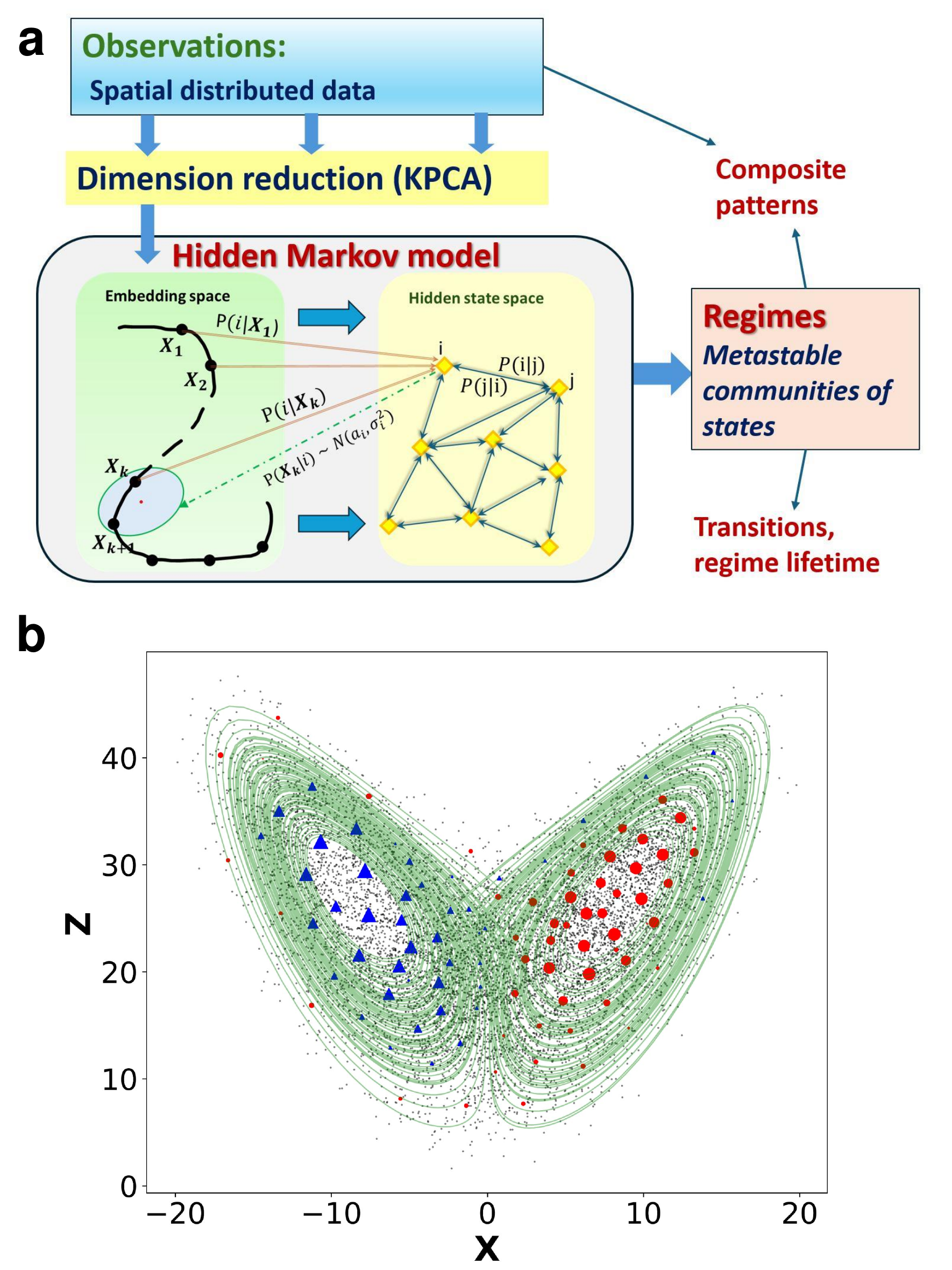}
	\caption{
    ({\bf a}) Schematic representation of the method for regime identification and analysis.
    ({\bf b}) Illustration of the method with the Lorenz model, Eq. (~\ref{lorenz}), showing the metastable states within the ($x,z$) phase space. The cloud of points (black dots) represent the "observations", whereas the filled circles and triangles represent the centers of the HMM hidden states (vectors $\mathbf a_J$ in Eq.~\ref{emissions}) for regimes 1 and 2 respectively. The size of these symbols is proportional to the stability measure $|w_{1i}|$, see Sec.~\ref{sec:pers_reg}. The green line represents a fragment of the trajectory on the attractor of the deterministic Lorenz system corresponding to $\sigma=0$.  
    }     
	\label{lor_scheme}
\end{figure}

Here we demonstrate the identification method using a low-order nonlinear chaotic dynamical system. The system is the classical Lorenz \cite{Lorenz1963} model with stochastic forcing:
\begin{equation}
\label{lorenz}
\begin{aligned}
& \dot{x}=10(y-x)+\sigma\xi_x\\
& \dot{y}=x(28-z)-y+\sigma\xi_y\\
& \dot{z}=xy-\frac{8}{3}z+\sigma\xi_z.
\end{aligned}
\end{equation}
Here $\sigma\xi$ are Gaussian uncorrelated processes with variance $\sigma^2=20$. From a 10,000-length trajectory generated by this system with time step 0.1, we consider the time series of $x$ and $z$ as our ``observations'' shown in Fig. ~\ref{lor_scheme}b. 
The (unforced) Lorenz system has 3 equilibria; an unstable center at the origin  and two metastable equilibria located at the centers of the butterfly wings. The high density around the origin 
(Fig. ~\ref{lor_scheme}b) is not genuine, and is partly due to the fact that this region corresponds to transition path between the two metastable equilibria, and partly to the stochastic forcing.
Next, a HMM has been constructed using these time series.  
As it can be seen from Fig.~\ref{lor_scheme}b, the obtained two communities of HMM hidden states represent well the metastable regions, and the most stable hidden states (see Sec.~\ref{sec:pers_reg}) lie close to, and around their centers.            
%%%%%%%%%%%%%%%%%%%%%%%%%%%%%%%%%%%%%%%%%%%%%%%%%%%%%%%%%%%%%%%%%%%%%%%%%%%%5

\subsection{Probabilistic analysis of regimes}

The obtained block structure of the matrix $\mathbf Q$ enables the calculation of reduced matrices $\overline{\mathbf Q^L}$ with elements defined by:
\begin{equation}
\label{transprob}
\overline{Q^L_{kl}}=\left[\sum\limits_{j\in A_l}{\pi_j}\right]^{-1}\sum\limits_{i\in A_k,j\in A_l}{Q^L_{ij}\pi_j}.
\end{equation}
The columns of this matrix represent the probability distributions of transitions between different regime states and allow us to predict the preferred transition path from any current regime. In particular, we can estimate the expected lifetime $\tau_k$ of the $k'th$ regime:
\begin{equation}
\label{exptime}
\tau_k=\overline{Q^1_{kk}}/(1-\overline{Q^1_{kk}})+1.
\end{equation} 
Note that the term $(\tau_k-1)$ coincides with the mean of the geometric probability distribution 
\begin{equation}
\label{geometric}
   P(\tau)=q^\tau (1-q),
\end{equation}
which represents the probability of a continuous series of length $\tau$ for an event with probability $q$.       

When dealing with real high-dimensional data, as the case here, one practically maps the observations $\mathbf Y$ onto a lower dimensional space $\mathbf X$, for HMM application, hence possibly compromising the inverse mapping onto the physical space. This is the case of
many nonlinear projections like kernel principal components (see the next section), diffusion maps, etc., for which we cannot derive an exact image of the hidden states and their groups in  the data space.
     
Here, to get anomalies associated with states $A_k$, we compute a composite pattern $\left<\mathbf Y|A_k\right>$, that is the mean of observed value $\mathbf Y$ given $A_k$ via     the obtained estimates of $P(J=j|\mathbf Y_t)$, i.e. probability that the observation $\mathbf Y$ at time $t$ belongs to the hidden state $J=j$ (see sec. 1.1). 
For each $\mathbf Y_t$, the regime occupation probability is first calculated:
\begin{equation}
\label{rprob}
P(A_k|\mathbf Y_t)=\sum\limits_{j\in A_k}{P(J=j|\mathbf Y_t)}, 
\end{equation}
which is then used to estimate the composite pattern:
\begin{equation}
\label{composite}
\left<\mathbf Y|A_k\right> = \frac{\sum\limits_t{\mathbf Y_tP(A_k|\mathbf Y_t)}}{\sum\limits_t{P(A_k|\mathbf Y_t)}}.
\end{equation}
Note that the community $A_k$, Eq. (~\ref{rprob}), can be used wholly or replaced by a subset corresponding, for example, to the most stable states (with the largest $|{w_1}_i|$). The occupation probability, Eq. (~\ref{rprob}), can also be used to derive other dynamical characteristics linked, for example, to long-term behavior such as mean number of days $D_k$  within $A_k$, in a given winter $T$:
\begin{equation}
\label{days}
D_k (T) = %\frac{1}{N_T}
\sum\limits_{t\in T}{P(A_k|\mathbf Y_t)}.
\end{equation}

\subsection{Kernel PCA}
\label{kpca}

To take into consideration the high-dimensional complex nonlinear structure of the atmospheric dynamical system, with possible nonlinear manifold of the dynamics an enable an efficient application of HMM, we apply nonlinear dimension reduction method based on kernel principal components (KPCs). Linear methods, such as conventional    EOFs, are unable to identify a nonlinear        low-dimensional embedding   of the system (e.g.\cite{HannachiIqbal2019a}).

Kernel principal component analysis (KPCA) has been shown \cite{Mukhin22} to efficiently   separate families of recurrent patterns of the mid-latitude atmospheric circulation and identify low-dimensional nonlinear manifold\cite{HannachiIqbal2019a}. KPCA belongs to the class of spectral methods and is based on embedding the data into a higher-, e.g. infinite-dimensional space where the system most likely becomes linear, and nonlinear manifolds and nonlinear features become linearly identifiable\cite{HannachiIqbal2019a}.  It allows implicit nonlinear mapping from the initial space by means of redefining the dot product, that tends to bring similar states closer together. Such a transformation makes it possible to embed the classes of similar states into low-dimensional subspace of a few leading scalar components. The measure of similarity of two states can be defined in the form of a Gaussian kernel
\begin{equation}
\label{kernel}
K(\mathbf Y_i,\mathbf Y_j)= \exp\left[{-\frac{d^2(\mathbf Y_i,\mathbf Y_j)}{2\sigma^2}}\right].
\end{equation} 
Here $d()$ is a distance between the states, which should be properly selected depending on the problem. Focusing on similarity of the pattern's structures/shapes rather than 
their amplitudes, we use Euclidean distance between the normalized vectors:
\begin{equation}\label{distance}
d(\mathbf x_i,\mathbf x_j)=\left\lVert\frac{\mathbf x_i}{\lVert\mathbf x_i\rVert}-\frac{\mathbf x_j}{\lVert\mathbf x_j\rVert}\right\rVert,    
\end{equation} 
where the norm of $\mathbf x$,  $\lVert\mathbf x\rVert=(\mathbf x^T\Lambda\mathbf x)^\frac{1}{2}$, is based on    a weight (diagonal) matrix $\Lambda$ to reflect the uneven density of grid nodes especially near the poles. Typically, the weight matrix depends on the latitude  $\theta_i$ of the $i'th$ grid cell. For example, if the focus is on %the polar region 
the global grid
then $\Lambda_{ii} =\cos \theta_i$, but if the focus is on midlatitude large scale dynamics then one uses (see the work \cite{Mukhin22} for details):
\begin{equation}
\label{weight}
\Lambda_{ii}=\cos{\theta_i}\sin^2\theta_i.    
\end{equation}

An eigenanalysis of the kernel matrix $K_{ij}=K(\mathbf Y_i,\mathbf Y_j)$ is then conducted after some kind of centering \cite{HannachiIqbal2019a} 
\begin{equation}
\label{centering}  
\mathbf {K_c}=\mathbf{C\cdot K\cdot C},
\end{equation}
with $\mathbf{C}=\mathbf{I}-\frac{1}{N}\mathbf{1}$,
$\mathbf I$ and $\mathbf 1$ are respectively the identity 
matrix and the matrix of ones. %This centering
The result of this centering operation 
can be interpreted as a dot product matrix calculated from a sample with zero mean.
The eigenelements of $\mathbf K_c$ are then derived, yielding our kernel PCs used in the HMM.  

We point out here that the hyperparameter $\sigma$ in Eq. (~\ref{kernel}) entails the degree of nonlinearity of the mapping. For instance, the smaller $\sigma$ is the better the local proximity preservation gets, that is nearby states in the input space map onto nearby states in the feature space. This is an attractive feature which allows easy cluster separation. Furthermore, when $\sigma$ is large it can be seen via Taylor expansion that for centered time series $\mathbf Y$ we have
${\mathbf {K_c}}_{ij} = \frac{1}{\lVert\mathbf Y_i\rVert\lVert\mathbf Y_j\rVert}\mathbf Y_i^T\Lambda\mathbf Y_j + o(\sigma^{-2})$, yielding similar results to conventional linear principal component analysis. 

The entire method described in this section is schematically represented in Fig.~\ref{lor_scheme}a. On the dimension reduction step the observed high-dimensional data vectors are mapped onto several KPCs, making thus a low-dimensional state space, which is used to separate classes of similar states. The HMM is then constructed and trained in this space, and communities of hidden states associated with metastable circulation regimes identified. Finally, dynamical and statistical characteristics of the circulation regimes are investigated along with association to large-scale teleconnections.

%%%%%%%%%%%%%%%%%%%%%%%%%%%%%%%%%%%%%%%%%%%%%%%%%%%%%%%%%%%%%%%%%%%%%%%%%%%%%%%
\begin{figure}[ht]
	\centering 
	\includegraphics[width=1.\textwidth]{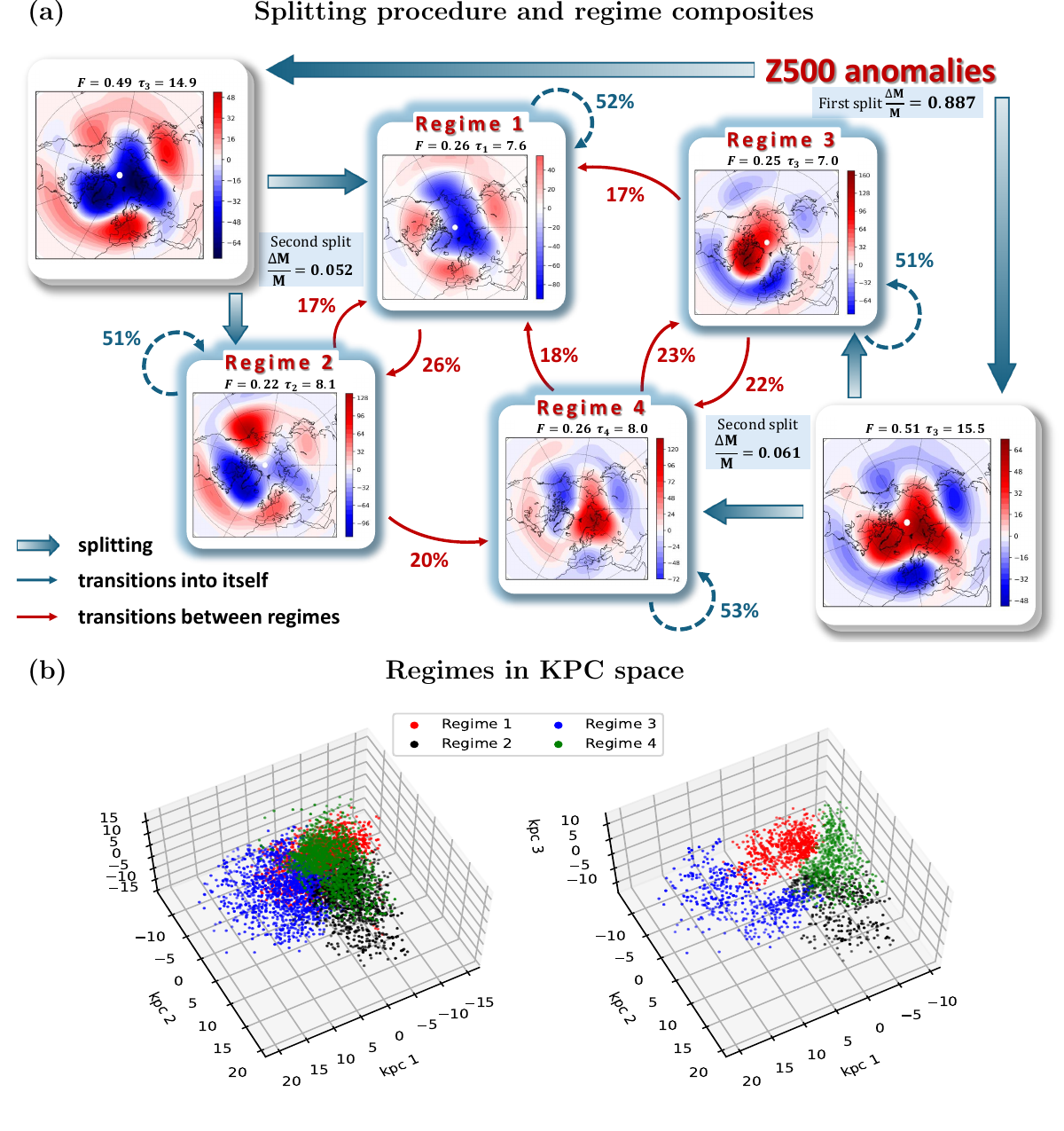}
	\caption{Circulation regime representation in data and KPC spaces. {\bf (a)} Composite patterns of communities obtained with $L=7$ days. Each pattern is calculated as a soft average of observed Z500 anomalies over a community of hidden states according to Eq.~\ref{composite}. Parent and child patterns of emerging communities during the iterative splitting process (see text) are shown by thick arrows. For each split, the relative increment $\frac{\Delta M}{M}$ of modularity is indicated, where $M$ is the final modularity after process completion. The four resulting communities are highlighted in the center of the diagram. Thin arrows indicate probabilities of transitions between regimes (red) as  well as self transitions (blue). Additionally, for each community, the fraction of residence $F$ (days) as   well as its mean lifetime $\tau$ (see Eq.~\ref{exptime}) are indicated. {\bf (b)} States of the system within the three leading KPCs color-coded according to their regime assignment showing points with assignment probability (see Eq.~\ref{rprob}) greater than 0.9 (left panel) and points belonging to 20\% of the most stable hidden states, see Sec.~\ref{sec:pers_reg}, (right panel).              
		}
	\label{fig_2}
\end{figure}

\section{Results}
\label{sec:results} 
\subsection{Circulation regime identification} 
We have, first, applied the method to simulation from a simplified system of atmospheric dynamics, namely, 
a 3-layer quasi-geostrophic (QG3) model \cite{Marshall1992} and check stability of the method. 
The results of this experiment are presented in the Supplementary Material. 
In this section we discuss the application to daily geopotential heights at 500-hPa pressure level (hereinafter Z500) in winter seasons over the mid-latitude of the Northern Hemisphere.
The Z500 fields are taken from the National Center of Environmental Prediction/National Center of Atmospheric  Research (NCEP/NCAR) reanalysis dataset \cite{Kalnay1996}. This is a gridded time series with                      
$2.5^o \times  2.5^o$  latitude-longitude resolution, that covers the period 1950--2023.  We focus here on the Northern Hemisphere mid-latitudes by considering latitudes north of 30N. The annual cycle is removed by subtracting the daily seasonal means over the whole period smoothed in time with a Gaussian window with the standard deviation of 15 days. This smoothing suppresses the day-to-day noise in the resulting annual cycle while retaining the intra-annual seasonal structure. And only winter (December--January--February) segments are taken, making a sample size of 6588 days, between winter 1950-51 and winter 2022-23. 

Relying on  previous findings \cite{Mukhin22}, for constructing of an HMM, we use a space of three leading KPCs, that provides good separating groups of similar states.    The HMM that has been derived from the described dataset has optimal number of hidden states equal to 50 (see Supplementary Material). 
We decomposed the obtained $50 \times 50$ transition matrix into metastable blocks for different powers $L$ determining the time steps of the analyzed evolution operator (see Sec.~\ref{sec:timestep}). Only significant splits of communities at the  1\% level were accepted during the splitting procedure. An example of how the states in the used KPC projection are arranged to the detected communities is presented in Fig.~\ref{fig_2}b, for the whole observed sample as well as for points attributed to the most stable hidden states within each circulation regime. It is seen that the communities are clearly separated in the KPC space, especially the most stable ones that form the cores of the regimes.      

\subsection{Structure of regimes and weather impact}

Figure ~\ref{fig_2} illustrates the splitting process of the communities for $L=7$. 
Other values of $L$ have been investigated and are presented in figures S3-S9 of the supplementary material.  These figures show the composite patterns of the obtained communities calculated using Eq. ~\ref{composite}. For spatio-temporal resolution $L \le 9$ days, the method yields consistently 4 highly significant communities. We notice, however, that the structure of the obtained circulation regimes with $L \le 5$ and $5 < L \le 9$ are somehow not quite similar. This dissimilarity can be explained by the existence of many short-lived recurrent fluctuations that corrupt the results at low $L$ which  cannot be correctly resolved due to lack of statistics. This is confirmed by examining an experiment by dropping the significance test and the threshold of the modularity increment $\Delta M$. We found (see, for example, supplementary figure S4) that, in case of low values of $L$ we get many more communities which vary with $L$, but for $L>5$ such spurious communities are filtered out and we obtain a few communities. 
Increasing the power $L$ will eventually lead to degeneracy of the transition matrix.
For instance, with $L>9$ the degeneracy first starts by merging regimes 3 and 4 yielding a broader community. With $L=11$ only two communities are obtained, which remain for larger values,  corresponding to an indivisible matrix. 
As a result, this consistently suggests that reliable and statistically significant persistent communities are detected for $5 < L \leq 9$, and, since the set of patterns  for this range of values are almost identical, we hereinafter discuss the results  corresponding to $L=7$.  

The first split of the initial community of all hidden states provides about 89\% of the relative community increment, and leading eventually to four communities (see Fig.~\ref{fig_2}). The communities resulting from the first split represent    opposing Z500 anomalies over the Arctic, Greenland, with anomalies in Eastern Europe and Far Eastern Siberia. Roughly speaking, these two communities reflect a Rossby wave structure with zonal wavenumber 4, and project respectively on states of intensified and weakened westerly flow. The following (and final) split yields smaller, but still significant, contribution to the modularity, providing identification of qualitatively different families of circulation patterns within the basic communities, hereinafter referred to as regimes 1 -- 4. 

\begin{figure}[ht]
	\centering 
    \includegraphics[width=1.\textwidth]{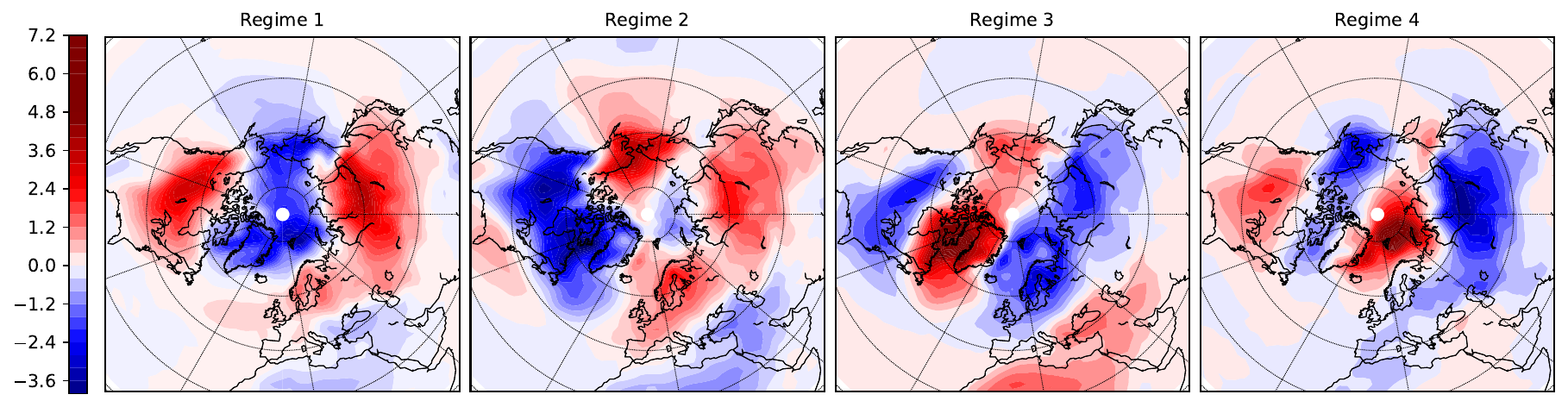}
	\caption{ 
    Composite of surface air temperature anomalies based on the resulting  Z500 circulation regimes.            
		}
	\label{fig:sat}
\end{figure}

These communities, regarded as metastable states in the low-frequency range, have longer time scales than typical synoptic systems and can therefore affect surface conditions over periods of time longer than synoptic time scales. To investigate the effect on surface weather we consider NCEP/NCAR daily surface air temperature (SAT) anomalies at 0.995 sigma level. The anomalies are computed in the same way as with the Z500 anomalies (see sec. 3.1). The regime occupation probabilities obtained for Z500 anomalies, Eq. ~\ref{rprob}, are used as weights to get the SAT composites patterns associated with the four circulation regimes (e.g., Eq.~\ref{composite}). The obtained SAT composites are shown in Fig.~\ref{fig:sat}.

Pattern 1 (Fig.~\ref{fig_2}), projects strongly on the positive phase of the Arctic Oscillation (AO) characterized by lower than normal pressure over the Arctic ocean. The  negative anomalies over the Arctic extends occasionally to reach the North Pacific and eastern Europe resulting in a stronger, than normal, jet stream there bringing weather systems from the west, e.g. loaded with rainfall. The second circulation regime Regime 2 combines features of the positive phase of the North Atlantic Oscillation (NAO), manifested as a deepened northern-subtropical Atlantic dipole in atmospheric pressure, and the negative Pacific North America Pattern (PNA) with an area of high pressure over the northeastern Pacific. 
A negative meridional pressure gradient in the North Atlantic, for both regimes, strengthens the zonal westerly flow in the Euro-Atlantic region, which may bring warm and wet conditions from Europe to Siberia (see SAT composites in Fig.~\ref{fig:sat}). 
Over North America, the two patterns have opposite effects. Regime 1 causes a  strengthening of the zonal flow bringing warm and wet oceanic air to this region,
especially over North east US/east Canada and the Canadian Archipelago. However, the blocking high pressure anomalies in the northern Pacific Ocean in regime 2 contributes   to pumping arctic cold air to the region.

The remaining two circulation patterns 3 and 4 (Fig. ~\ref{fig_2}) emerge from the splitting of the community characterized by high pressure over the Arctic, i.e.
negative phase of AO.  Pattern 3 clearly projects on the negative NAO phase, which prevents moist and warm Atlantic air masses from reaching Europe and contributes to  cooling northern Eurasia. At the same time, the northwest Atlantic area, including Greenland and Labrador sea, are getting warmer than normal, in addition to warmer and wetter, than normal, Mediterranean. In the circulation pattern 4, anticyclonic anomalies are shifted eastward, centered over the Barents sea. This pattern contributes to cooling northern Siberia, but also warms the Eurasian basin of the Arctic Ocean. 

%%%%%%%%%%%%%%%%%%%%%%%%%%%%%%%%%%%%%%%%%%%%%%%%%%%%%%%%%%%%%%%%%%%%%%%%%%%%

\subsection{Transition probabilities, lifetimes and pattern diversity}

Another, yet important characteristic feature of the identified patterns is their lifetime and transitions. The transition probabilities between the regimes, i.e. the reduced matrix calculated via Eq.~\ref{rprob} with $L=7$ is 
\begin{equation}
\label{rtrans}
\overline{\mathbf Q^L}\approx\left(
\begin{matrix}
0.52 & 0.17 & 0.17 & 0.18 \\
0.26 & 0.51 & 0.1 & 0.06 \\
0.13 & 0.12 & 0.51 & 0.23 \\
0.09 & 0.2 & 0.22 & 0.53
\end{matrix}
\right)
\end{equation}  

Several points can be inferred from the above transition matrix. First and foremost we
can derive the probability distribution of the likelihood of the transition to the
next circulation regime.  
These distributions are clearly different from the uniform PDF, as it is seen from Fig.~\ref{fig_2} (thin arrows). Circulation regimes 3 and 4, for example, are more likely to interact with each other than with the remaining regimes. Both these circulation patterns concern similar blocking structures in the Euro-Atlantic region, and during the coldest Eurasian winters they often alternate in time (see Supplementary Animation 1).  The next likely transition, from these two patterns, is toward circulation pattern 1,  which, in turn, prefers to resolve into regime 2. However, the preferred path from the latter is regime 4, although the probability of switching to regime 1 is also high. These transitions lead to a preferred loop of circulation regime dynamics, namely,  $r1\to(r2\leftrightarrow r1)\to(r4\leftrightarrow r3)\to r1$.             

\begin{figure}[ht]
	\centering 
	\includegraphics[width=1.\textwidth]{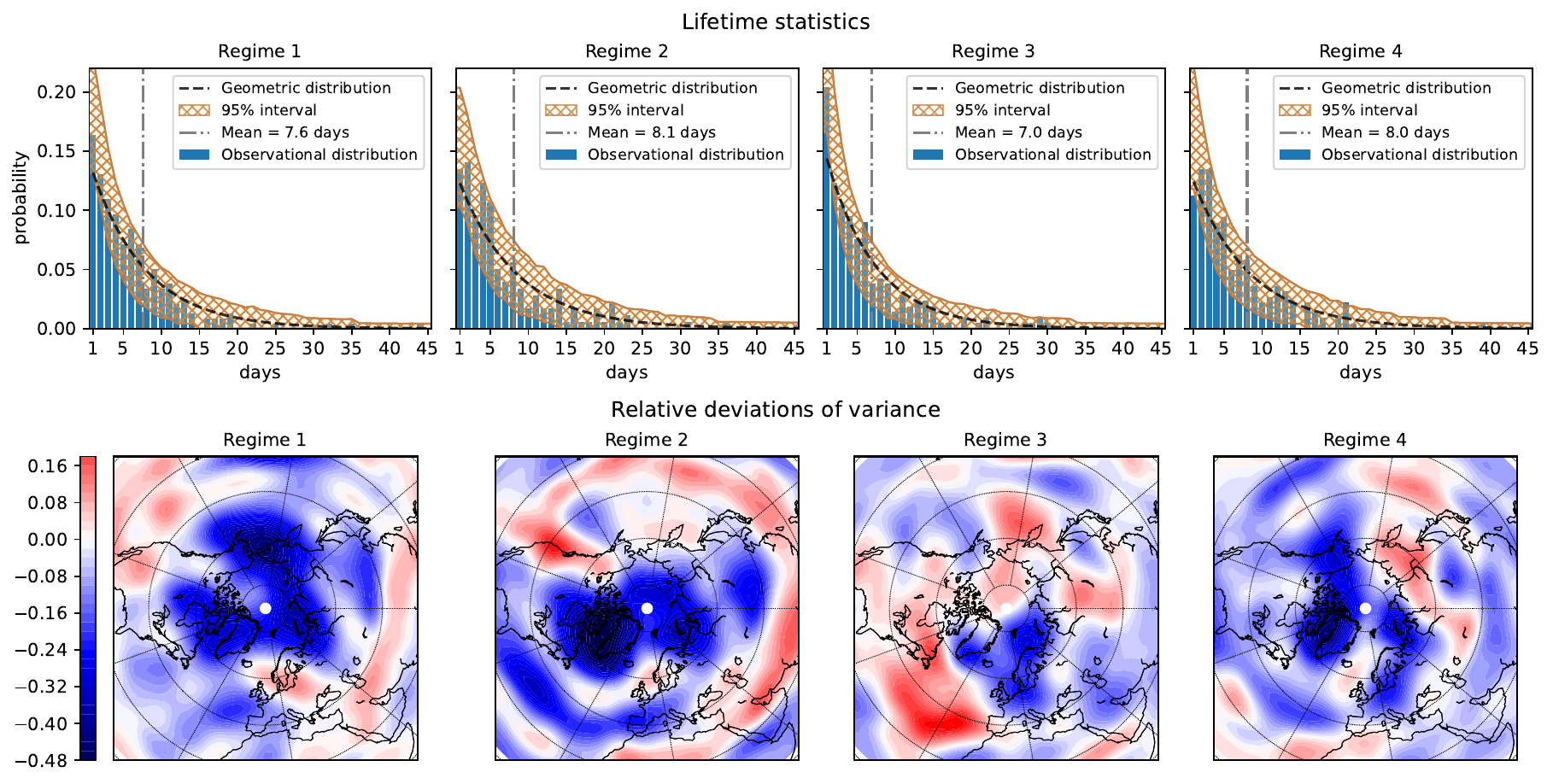}
	\caption{Properties of the regimes.
		{\bf Top panel}: distributions of regime lifetimes. Blue bars represent observational distribution, i.e. interval statistics of sequences of observations assigned to the regime with probability more than 0.95. Orange shade shows 95\% confidence interval of an ensemble of 10,000 realizations generated by the HMM.
		Geometric distribution Eq.~\ref{geometric} with a parameter $q$ equal to the probability of a regime to transit into itself for a one-day interval (i.e. the diagonal element of the one-day reduced transition matrix), as well as its mean, are indicated by dashed and dash-dotted line, respectively.
		{\bf Bottom panel}: relative Z500 variance anomalies within a regime with respect to the sample variances (see Eq.~\ref{relvar}).            
		}
	\label{fig_3}
\end{figure}

The probabilities of self transitions, given by the diagonal elements of the transition matrix, $\overline{\mathbf Q^L}$, Eq.~\ref{rtrans}, can be used to estimate a posteriori the distribution of regimes lifetime via Eq.~\ref{geometric}. At the same time we can use the HMM to generate an ensemble of surrogate time series mimicking the observations. These surrogates can be used to derive various statistics, such as regime lifetime distribution, and can be used in hypothesis testing. Figure ~\ref{fig_3} shows the obtained lifetime distribution compared to that from the observations for each of the circulation patterns. Overall the figure shows comparable values of the expected lifetimes of all circulation regimes, ranging from 7 to 8 days, though an anomalous maximum,  of 25-30 days, can be noted in the observational distribution of circulation pattern 3.  In the meantime, Fig.~\ref{fig_3} also shows that such rare events fit well into the model statistics and cannot be rejected based on the quasi-geometric model distribution.  We caution, however, that the available statistics are not sufficient to cover the tail of the observational distribution well because of the sample size, and there is no decisive evidence to suggest that regime 3 tends to live systematically longer than other regimes.  

To investigate the similarity of the atmospheric circulation within the patterns we show  in Fig.~\ref{fig_3} the relative variance of observations given a specific regime $k$:
\begin{equation}
\label{relvar}
\mathbf R_k=\frac{\mathrm\Omega_k-\overline{\mathrm\Omega}}{\overline{\mathrm\Omega}},
\end{equation}
where $\mathrm\Omega_k=\left<\mathbf Y^2|A_k\right>-\left<\mathbf Y|A_k\right>^2$ is the variance of the $k'th$ regime (the operator $\left<\right>$ is defined as in Eq.~\ref{composite}), and $\overline{\mathrm\Omega}$ is the total sample variance of the observations. The quantity $\mathbf R_k$ provides a measure of intra-regime variance.
Overall, we see low intra-regime variance in areas of high values of anomaly composites  for regimes 1, 2, and 4 (shown in Fig.~\ref{fig_2}). But this is not the case for regime 3, which shows small variance in the Euro-Atlantic region, while in the subtropics the variance significantly exceeds the total sample variance.  This circulation regime projects onto $-$NAO for which the blocking high around Greenland is known by its high persistence \cite{woollings2010}. But the subtropical low
%%%%%%%%%%%%%%%%%%%%%%%%%%%%%%%%%%%%%%%%%%%%%%%%%%%%%%%%%%%%%%%%%%%%%%%%%%%%
%%%%%%%%%%%%%%%%%%%%%%%%%%%%%%%%%%%%%%%%%%%%%%%%%%%%%%%%%%%%%%%%%%%%%%%%%%%%
pressure system is not as persistent explaining the larger variance there. It is also possible that the weakening   of the pressure gradient between the polar region and the midlatitudes yields unstable disturbances that can travel equatorward reaching the subtropics and modify the flow   there (see Supplementary Animation for the winter 2009-2010 with dominating regime 3). A similar situation occurs for regime 2 over the North Pacific, where the blocking high leads to increased variances further south. 
 %

%%%%%%%%%%%%%%%%%%%%%%%%%%%%%%%%%%%%%%%%%%%%%%%%%%%%%%%%%%%%%%%%%%%%%%%%%%55

\subsection{Long-term dynamics of circulation regimes and teleconnection}
  
\begin{figure}[ht]
	\centering 
	\includegraphics[width=1.\textwidth]{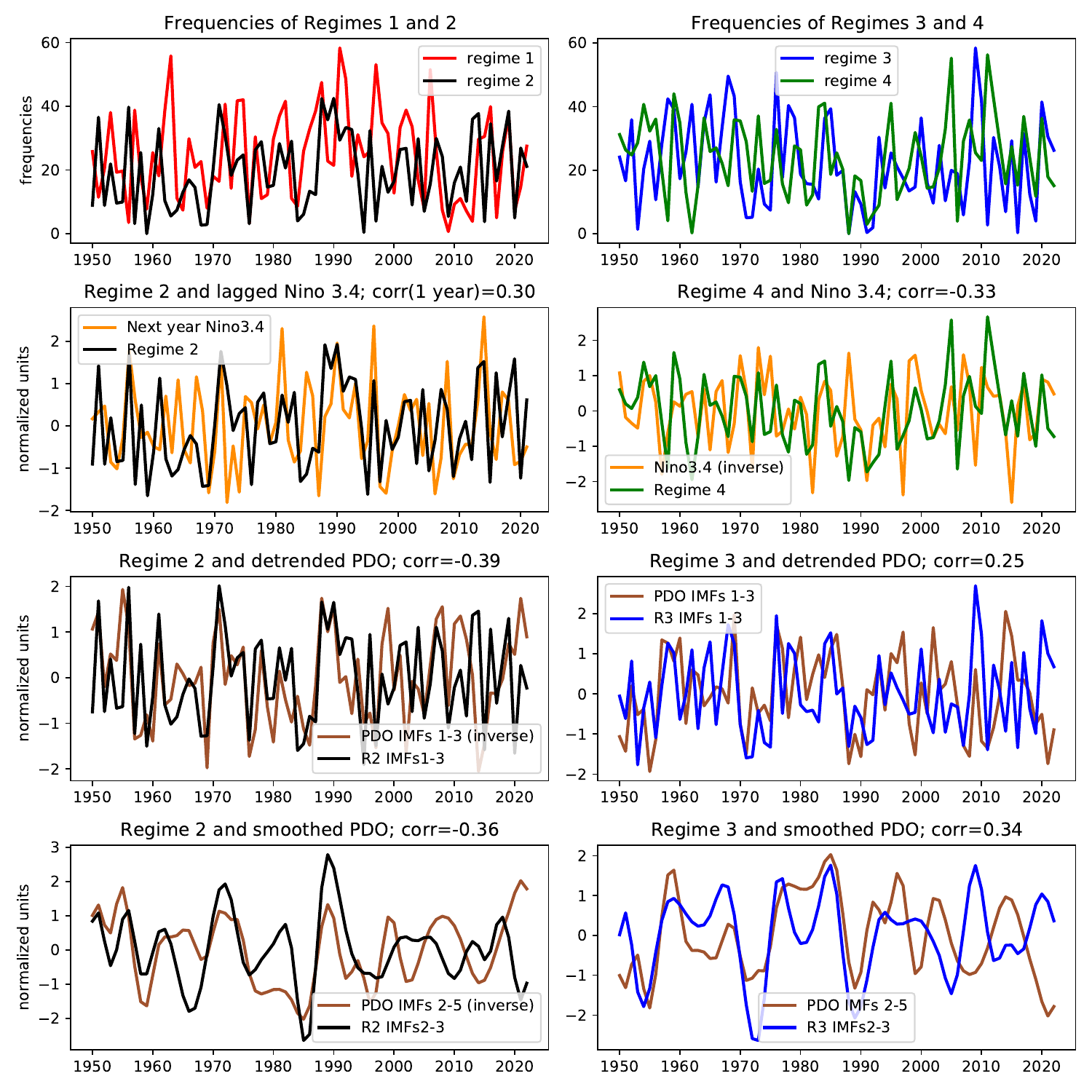}
	\caption{
    Interannual fluctuation of regime frequency and link to large-scale teleconnection.    
	{\bf Top row}: Frequency of the number of days in each regime during winter,  see Eq.~\ref{days}. 
    {\bf Second row}: Regimes 2 and 4 frequency and ENSO showing the Ni\~no 3.4 index with 1-yr lag (left) and (opposite) Ni\~no 3.4 (right). 
    {\bf Third row}: Regimes 2 and 3 frequency and detrended PDO index filtered using the first three IMFs. 
    {\bf Fourth row}: Regimes 2 and 3 frequency and smoothed PDO index by removing the first (short time-scale) IMF.
    When plotted together with regime 2 frequency, the PDO index is multiplied by $-1$. Correlation coefficients of the regime frequency with the (non-inverted) indices are also indicated. The year in the x-axis refers to that og January.  
    }
	\label{fig_1}
\end{figure}

One of the question of importance here is related to the characteristics and variability across scales of these patterns. Figure ~\ref{fig_1} shows the mean number of days per winter, i.e. regime frequencies, as in Eq.~\ref{days}. The figure exhibits pronounced interannual (2--4 year) and decadal variability. Note that we attempted to assemble the  time series into pairs based on the blocking action of the zonal flow in the Euro-Atlantic region resulting in either warm or cold winters over northern Eurasia (see Fig,~\ref{fig:sat}). Although cross-regime correlations appear quite low on interannual timescales, we observe clear decadal correlations between the "warm" regimes 1 and 2 as well as the "cold" regimes  3 and   4.  In particular, there is an approximate 10-year period of warmer than normal winters extending from the mid-80s to the mid-90s, a consequence of the abnormally high difference between the frequencies of "warm" and "cold" regimes. Opposite conditions arose in the 1960s and early 2010s. Examples of atmospheric behavior in two winters with opposite anomalies in the Euro-Atlantic region are presented in the Supplementary Animations, where it is shown that 2009-2010 and 1991-1992 winters are dominated by regimes 3 and 1, respectively. 

Detailed analysis of whether this dynamical behavior can be attributed to decadal climate modes such as Pacific decadal oscillation (PDO) or Atlantic multi-decadal oscillation (AMO), or whether it is a manifestation of internal atmospheric variability known to be a characteristic feature of mid-latitudes, goes beyond the present paper, and is subject of further research. Here we only report significant correlations of regime frequency with El-Ni\~no Southern Oscillation (ENSO) represented by the Ni\~no 3.4 index \cite{Rayner2003} and PDO index \cite{Newman2016}.

%%%%%%%%%%%%%%%%%%%%%%%%%%%%%%%%%%%%%%%%%%%%%%%%%%%%%%%%%%%%%%%%%%%%%%%%%%%%%%%%5

\begin{figure}[ht]
	\centering 
	\includegraphics[width=1.\textwidth]{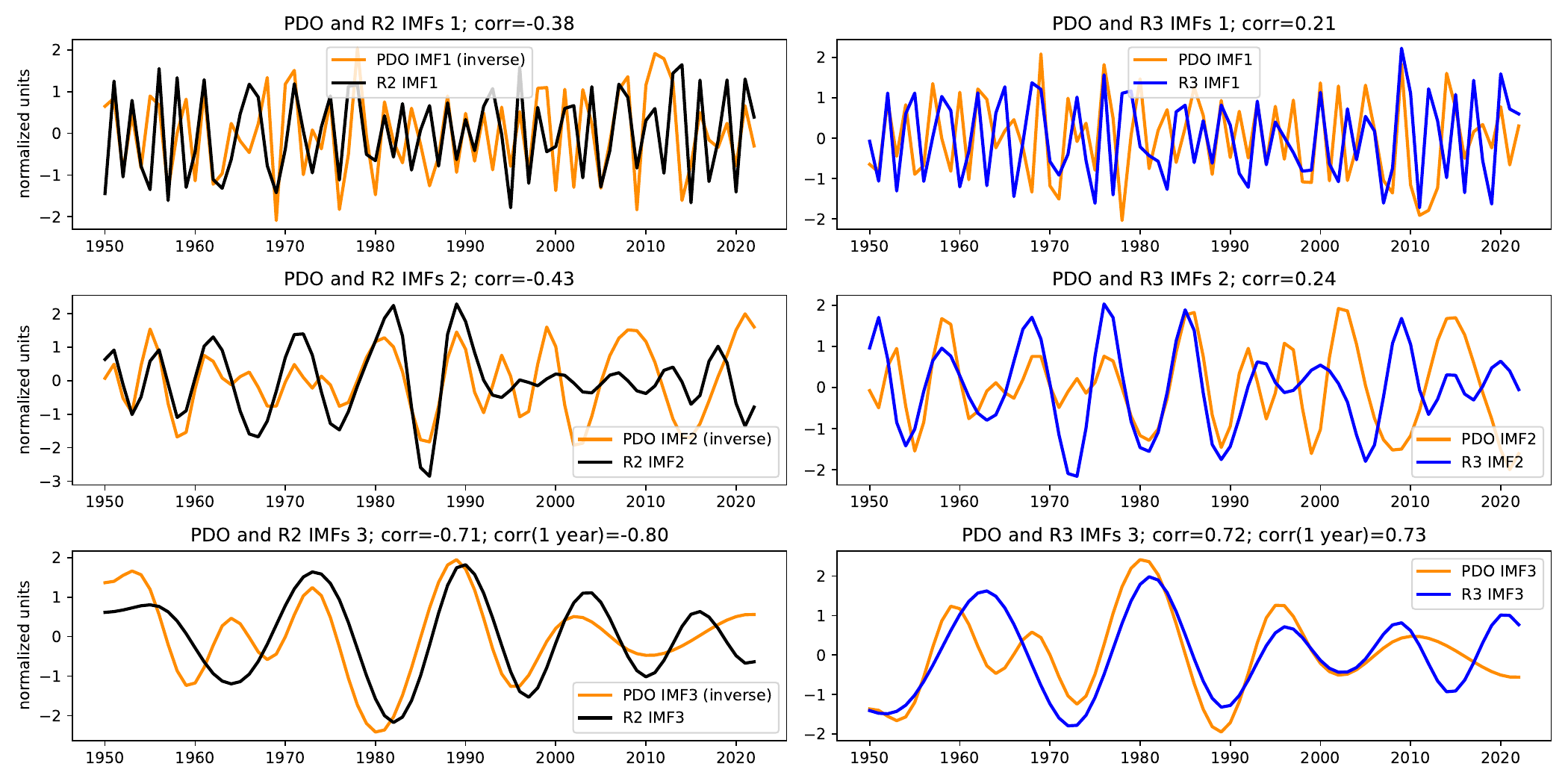}
	\caption{Leading three IMFs of PDO winter mean index (orange), regimes 2  (black) and regime 3 (blue) frequencies. The PDO IMFs time series in the   middle and bottom left panels are multiplied   by -1 (inverted) for  convenience. Correlation coefficients of regime IMFs with corresponding PDO  IMFs (non-inverted) are also indicated. In the bottom panels correlation of   PDO IMF3 with 1-year lagged IMFs 3 of regimes 2 and 3 are also shown. The year in the x-axis refers to that of January.  
    }
	\label{fig:imfs}
\end{figure}

ENSO is one of the dominant modes of climate variability bringing about a complex network of teleconnections worldwide \cite{alexander2002,Trenberth2019}. The Ni\~no 3.4 index defined as mean sea surface temperature in the central tropical pacific region (5S-5N and 170-120W) is widely used for ENSO monitoring and forecast \cite{Barnston2012}. Correlation analyses between the Ni\~no 3.4 index and regime frequency produce small but significant correlations with regimes 1,2 and 4 (see Fig.~\ref{fig_1} and \ref{fig:imfs}). Positive correlation with regime 1 is likely due to the influence of ENSO on the positive PNA-like anomaly \cite{Mo1986} in the North Pacific (see Fig.~\ref{fig_2}). Interestingly, there is a lagged influence   of regime 2 on ENSO, but with no significant contemporaneous correlation. This  indicates that negative PNA, which projects onto regime 2, impacts North Pacific Oscillation mode in the following months making it favorable for developing El-Ni\~no in the following season, see, e.g. \cite{Vimont2003b,Mukhin2021a}. Finally, there is a negative correlation between winter Ni\~no 3.4 and circulation regime 4. Because this regime is dominated by high pressure anomaly over north/eastern  Europe, its association with the culminating ENSO phase suggests an influence of the latter on weather in the Euro-Atlantic region.

The PDO, a dipole of SST pattern in the North Pacific, is driven by internal   feedbacks  from the complex interaction of coupled ocean-atmosphere system and by       tropical/extratropical forcing. This results in a wide spectrum of PDO including interannual, decadal, and longer timescales.
In an attempt to scrutinize the relation to teleconnections we filter out high-frequency variability and compare the obtained time series. We use Empirical Mode
Decomposition (EMD) \cite{Huang2008} -- a powerful data-driven approach for  extracting dominant intrinsic modes/components ranked by their characteristic timescales, referred to as Intrinsic Mode Functions (IMFs). These IMFs provide a consistent smoothing tool by discarding for example the fastest modes or by removing long-term trends.
We identified highly significant correlations of the PDO index with regimes 2 and 3 frequency both in the short time-scale band, dominated by 1-10 year periods, and in  the intermediate, 10-20 year, band (see Fig.~\ref{fig_1}). 
The highest correlation, around -0.4, is reached with the short time-scale component of the PDO index and regime 2 frequency time series, both represented by IMFs 1--3.   
This reflects a significant impact of PDO-induced SST anomalies in the eastern North Pacific on the mass field and geopotential height anomalies where regime 2 is pronounced. A significant positive, but lower, correlation is obtained with regime 3 frequency. 
For the intermediate time scale, we consider a smoothed PDO index by discarding the high-frequency IMF-1 and smooth the regime frequency time series using IMFs 2--3. Figure ~\ref{fig_1} shows highly significant negative and positive correlation 
with regimes 2 and 3, respectively. 
Since regime 3 is largely dominated by the NAO pattern, the obtained correlation suggests interactions between North Pacific and North Atlantic climate variability  on decadal time-scales. To highlight the effect of these timescales, 
Figure~\ref{fig:imfs} shows the correlations between PDO index and regimes 2 and 3 frequency using IMFs 1--3. Unexpectedly high correlations are obtained with IMF-3 of PDO and 1-year lag regimes frequency of the order -0.8 for regime 2 and 0.73 for regime 3.
Correlations based on IMFs 1 and 2 are also highly significant for regime 2 with  values -0.38 and -0.43 respectively. 
It is worth noting that the time period from the 70's to the 90's, identified as a positive PDO epoch \cite{Trenberth2019,Mukhin2018}, has the largest contribution to the obtained correlations.     

%%%%%%%%%%%%%%%%%%%%%%%%%%%%%%%%%%%%%%%%%%%%%%%%%%%%%%%%%%%%%%%%%%%%%%%%%%%%%

\section{Discussion}
\label{sec:discussion}

We present a novel methodology of data-driven detection of recurring atmospheric circulation patterns by considering these patterns as metastable communities of the nonlinear system dynamics. The  nonlinear formulation of the problem effectively narrows the class of recurrent states of the atmospheric flow into circulation regimes. This is based on building a dynamic stochastic model of the system evolution operator, allowing  one to determine areas in the state space of trajectory slowing down.  Data-driven nonlinear modeling is known to be a challenging inverse problem, and nonlinear optimization is needed, see e.g. \cite{Mukhin2019,Seleznev2022}.
Here we adopt a simplified representation of the system via a finite-state Markov model whose complexity, or resolution, depends on the extent of observational statistics. The approach recasts the nonlinear dynamics of the original system into a linear operator on a set of probability distributions, and where metastability can be captured via stochastic matrix analytic methods. 

The method uses HMM combined with nonlinear dimension reduction and community splitting based on graph  theory. Given the available sample size, a substantial part of our analysis is devoted to investigating the reliability of the results. First, each community splitting must pass a fairly stringent significance test before acceptance. Next, the degree $L$ of the transition matrix, which determines the model resolution, is adjusted, and the range corresponding to stable results is selected. This contributed to results reliability and helped yield roughly coherent metastable structures of the system dynamics. The method has been successfully applied to a low-order chaotic model and a 3-level quasi-geostrophic model, which is a simplified low-complexity atmospheric model of midlatitude dynamics. The method is then applied to the NCEP/NCAR Z500 over the NH during winters 1950-2023. Four (nonlinear) circulations patterns are identified. These patterns partially project onto NAO and PNA. Lifetime and likely transitions between these metastable patterns are then identified.

One important question that arises often in discussions is related to the connection between these patterns and "standard" modes of atmospheric teleconnections emerging in atmospheric LFV.  The typical patterns, such as NAO, PNA, and AO, etc., have historically been isolated empirically by studying regional dynamical processes and by their influence on weather systems, and then confirmed via linear correlation analysis \cite{Wallace1981}.  However, from a global perspective, all these modes are manifestation of a more general evolution  of the planetary-scale atmospheric circulation involving many nonlinear and complex processes, including lower boundary conditions, nonlinear wave-wave interaction, and complex forcing patterns.  The dynamical feature of atmospheric circulation manifests itself partly in those linear modes,  but the full picture needs a deeper investigation into the nonlinear structure of the system trajectory within its state space. The metastable regimes constitute the backbone elements of the dynamics of this global system and  are of paramount importance for understanding the dynamics of all regionally significant large-scale processes,    and helps toward extending the chaotic limits of predictability. In particular, in Sec.~\ref{sec:results} we show that each of the obtained regimes contributes to those well-known teleconnections, and can provide suggestions for predictability of various atmospheric flows.     The presented method allows to extract areas of metastability, but also gives a consistent probabilistic assessment of their evolution via the transition matrix.  

The method presented here can provide potential for subseasonal-to-seasonal (S2S) prediction. One pivotal direction of the method extension consists in adopting it for long-term regime prediction, including inter-seasonal forecasting. This requires a model that evolves continuously across the  year involving the interaction between the circulation regimes and the annual cycle. To expand the current stationary HMM, our future model extension will include forcing by nonstationary teleconnection and includes the annual variability.

%\clearpage

\section*{Acknowledgements}
The work is supported by the Russian Science Foundation (grant \#22-12-00388). The implementation and testing the algorithms is supported by the Scientific and Educational Mathematical Center (Agreement with the Ministry of Science and Higher Education of the Russian Federation No. 075-02-2020-1632).

\section*{Author contributions statement}
D.M. and A.H. conceived the project and wrote the manuscript. D.M. and R.S. implemented the methods and performed calculations. All authors analyzed the results and reviewed the manuscript. 

\section*{Additional information}

\textbf{Competing interests}

The authors declare no competing interests. 

%\bibliography{refs}

\end{document}


% \baselineskip24pt

% % Make the title.
\maketitle
% \begin{center}
% \center{\LARGE Supplementary materials for}
% \center{\Large Inferring the Middle Pleistocene Transition Mechanism from Data-Driven Models}
% \center{\large Dmitry Mukhin, Andrey Gavrilov, Evgeny Loskutov, Juergen Kurths, Alexander Feigin}
% \center{\large Correspondence to: mukhin@ipfran.ru}
% \end{center}

\section*{Supplementary text}

\paragraph*{Significance testing}

According to the method described in Sec. 2.2 of the main article, a split of a community $A$ of HMM hidden states can be realized only if it provides positive increment $\Delta M_A$ of the modularity function $M$. However, to ensure this increment is significant, we  need to reject the null hypothesis that the split yields the same increment by chance, i.e. for a sample taken from a random process without metastable regimes. In order to make the test stronger, we require the random process to have     the same invariant distribution as that of the original HMM process and with similar spectral properties on short timescales. We use an ensemble approach to formulate the null hypothesis by using samples generated by means of the original HMM operator, but with shuffled time blocks. Using the original HMM as a generator of surrogates guarantees preserving the invariant measure, whereas block shuffling destroys autocorrelations crucial for persistent regimes. 

The important hyperparameter of this procedure is block length, which determines autocorrelation properties of surrogates. For example, primitive shuffling with unit blocks corresponds to testing against a white noise process. In order to make the spectrum of the surrogates closer to the spectrum of the original HMM process at short time scales, we use distribution of block lengths derived from the HMM transition matrix decomposition. The evolution of an initial distribution ${\bf q}$ over discrete time $t$ under the iterated Markov operator $\mathbf Q^t$ can be decomposed using the eigenbasis of the stochastic matrix $\mathbf Q$ columns as
\begin{equation}
\mathbf Q^t\mathbf q=\pi+a_2\lambda_2^t\mathbf u_2+a_3\lambda_3^t\mathbf u_3+\dots,    
\end{equation}
where $1>\lambda_2>\lambda_3>\dots$ are eigenvalues of the stochastic matrix $\mathbf Q$, with respective right eigenvectors $\pi,\mathbf u_2,\mathbf u_3,\dots$ where $\pi$ is the stationary distribution, and $a_i$ is a coordinate of $\mathbf q$ along the eigenvector       $\mathbf u_i$. This decomposition shows that during convergence of an initial distribution    $\mathbf q$ to the stationary distribution $\pi$, different components of $\mathbf q$    vanish with different e-folding times  $T_i=1/\log|\lambda_i|$. This spectrum                  $\Theta=\{T_2,T_3,\dots\}$ of characteristic times is then used as a distribution for the block lengths to generate surrogates.

Technically, a model time series is generated by iterating the Markov chain of hidden state probabilities inside the HMM and sampling the emission PDFs at each iteration to obtain a state in the KPC space. To generate a shuffled time series, we randomly draw $T_i$ from $\Theta$, iterate the HMM $T_i$ times, and then randomize the Markov chain using new initial conditions from the stationary distribution $\pi$. A new value from $\Theta$ is then drawn and the HMM iterated  again, etc. The procedure continues until the required time series length is reached.

For the significance test we use an ensemble of 10,000 surrogate times series, and from   each of these we obtain a surrogate transition probability matrix, given the fixed emission PDF parameters taken from the original HMM. When splitting a current community $A$ (see Sec. 2.2) into two new communities in the original HMM, we also measure the modularity increment   $\Delta M$ from this split in each surrogate. The obtained ensemble of increments provides  us with a null hypothesis distribution of $\Delta M$, which is used to decide on the acceptance/rejection of the split. A $\Delta M$ below a significance level 1\% is treated   as non significant and the corresponding split rejected.    

\paragraph*{Regimes of QG3 model}

Quasi-geostrophic (QG) models of the atmosphere are widely used for simulating realistic   mid-latitude atmosphere behavior, see e.g. \cite{Kondrashov2003, Mukhin2022,HannachiIqbal2019b, Seleznev2019}.  QG models, which are based on quasi-geostrophic potential vorticity conservation, demonstrate a rich spectrum of variability at different time scales and are competitive to intermediate and full general circulation 
models  regarding complexity and  dynamical features/processes. Here we use several time series generated by a three-level QG model (QG3) on the sphere \cite{Marshall1992} with realistic orography and surface boundary condition. Based on the equations of the potential vorticity at three (200, 500, and 800 hPa) pressure levels \cite{Marshall1992,Vannitsem1997,Corti1997}, the model is tuned to simulate winter atmospheric circulation in the extratropical hemisphere. Here we present results of HMM regime detection independently applied to four 10,000-day time series of the mid-level stream function anomalies, distributed in latitudes 36\textdegree N to 90\textdegree N with approximately $5.5 \times 5.5$ degree resolution. These non-overlapping time series are randomly taken from a very long (300,000 days) QG3-model run.

\begin{figure*}[ht]
\centering
\includegraphics[width=1.\textwidth]{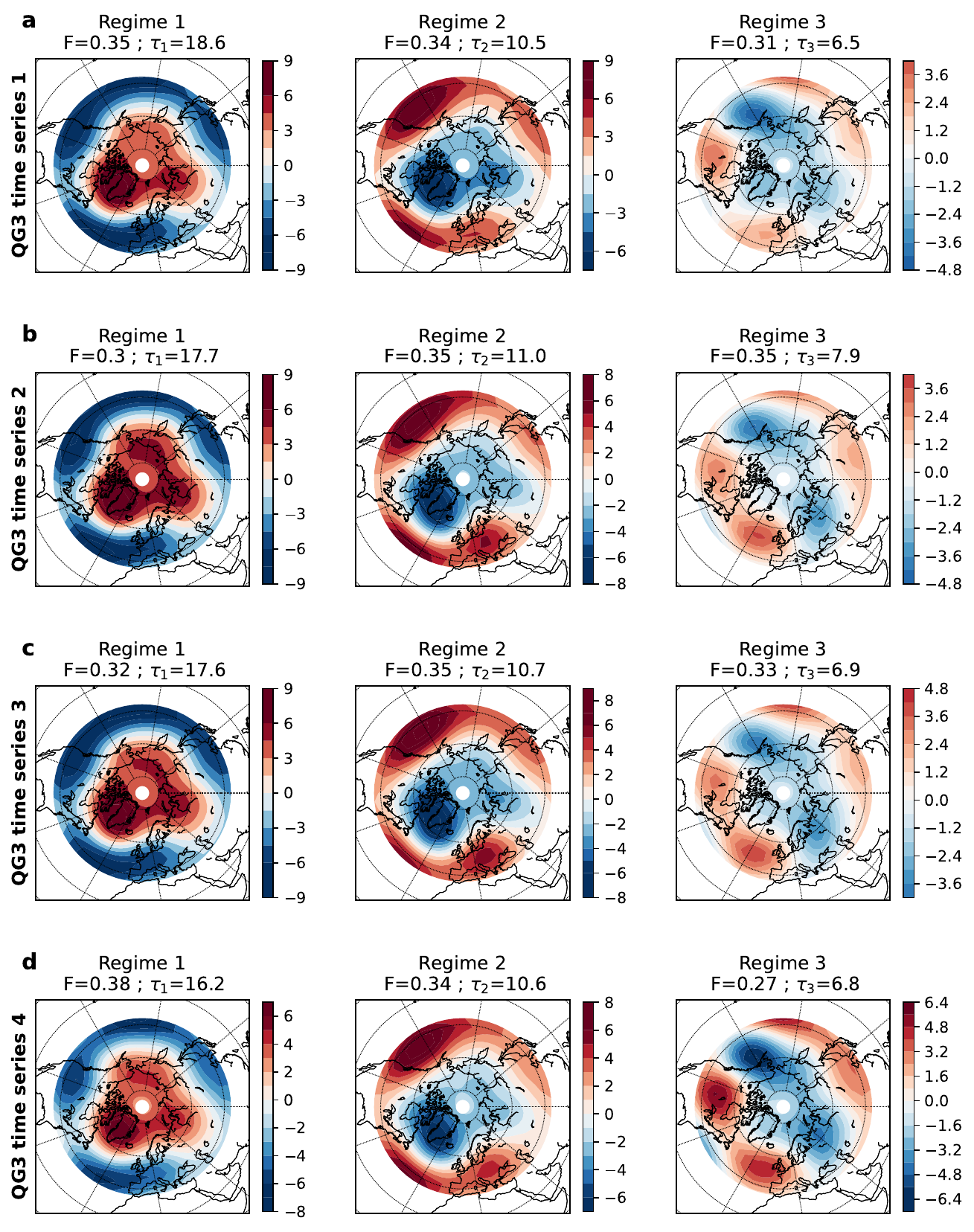}
\caption{Regime composites for QG3 model times series. For each circulation regime the    mean anomalies of the stream function from the middle level of the model are shown. The  fraction of the number of days $F$ that the system spends in each regime as well as regime mean lifetime $\tau$ are indicated. }
\label{composites}
\end{figure*}

The analysis, described in Secs.~3.1-3.2 of the main article, applied to each of the four  QG3 independent time series reveal 3 regimes, which are well-reproduced at HMM time steps $7<L<11$ days. Composite patterns of the obtained regimes shown in Fig.~\ref{composites}   are nearly identical for all the time series. The two most stable regimes 1 and 2 correspond to opposite, negative and positive phases of AO, respectively. Regime 1 is highly persistent, with mean lifetime of 16-18 days. Regime 2, which has mean lifetime 10-11 days, also has anomalies typical for NAO as well as high anomalies in North Pacific resembling the negative PNA pattern. The last short-lived regime 3 contains a combination of states with moderate anomalies; which can be interpreted as transient states between metastable regimes 1 and 2. It is worth noting that the characteristic timescales of QG3 model dynamics, including regime persistence, are longer than of observed behavior (see results with reanalysis data in Sec. 3). Also, its regime composition is much simpler as compared with reanalysis.              

\newpage

\section*{Supplementary figures}

\begin{figure*}[ht]
\centering
\includegraphics[width=1.\textwidth]{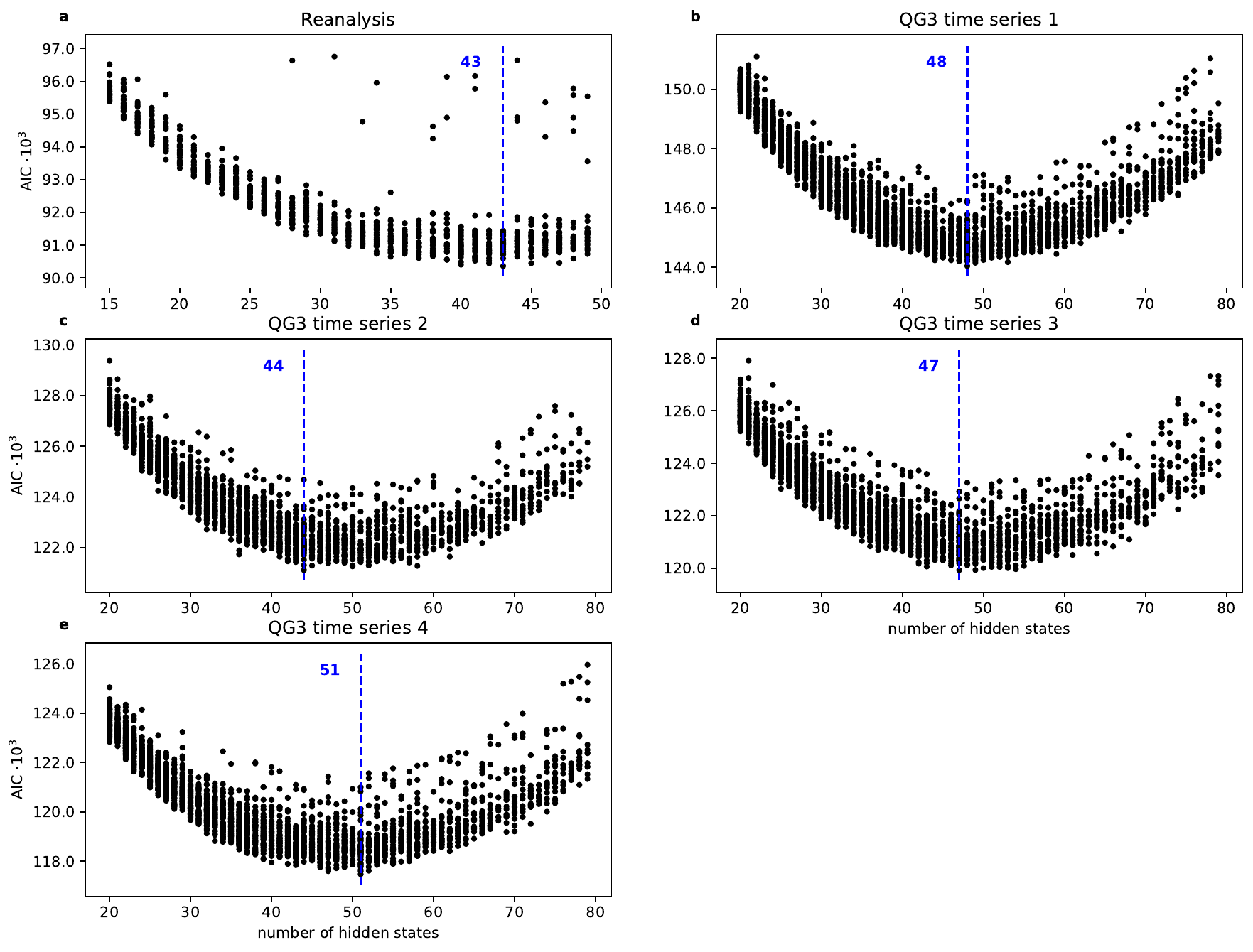}
\caption{Optimization of the number of hidden states by Akaike Information Criterion (AIC). AIC optimality is defined as $2K-\log(P(\mathbf Y|\mu,K))$, where K is the number of states and $P(\mathbf Y|\mu,K)$ is the likelihood of HMM with parameter $\mu$ relative to the time series $\mathbf Y$. This value is plotted vs. K for reanalysis Z500 data (see the main article) as well as for each of the four QG3 time series. For each value of K the HMM was trained many times from different initial parameters of emission PDFs. Finally, the optimal model was selected by selecting the optimal K (blue).}
\label{aic}
\end{figure*}

\newpage

\begin{figure*}[ht]
\centering
\includegraphics[width=1.\textwidth]{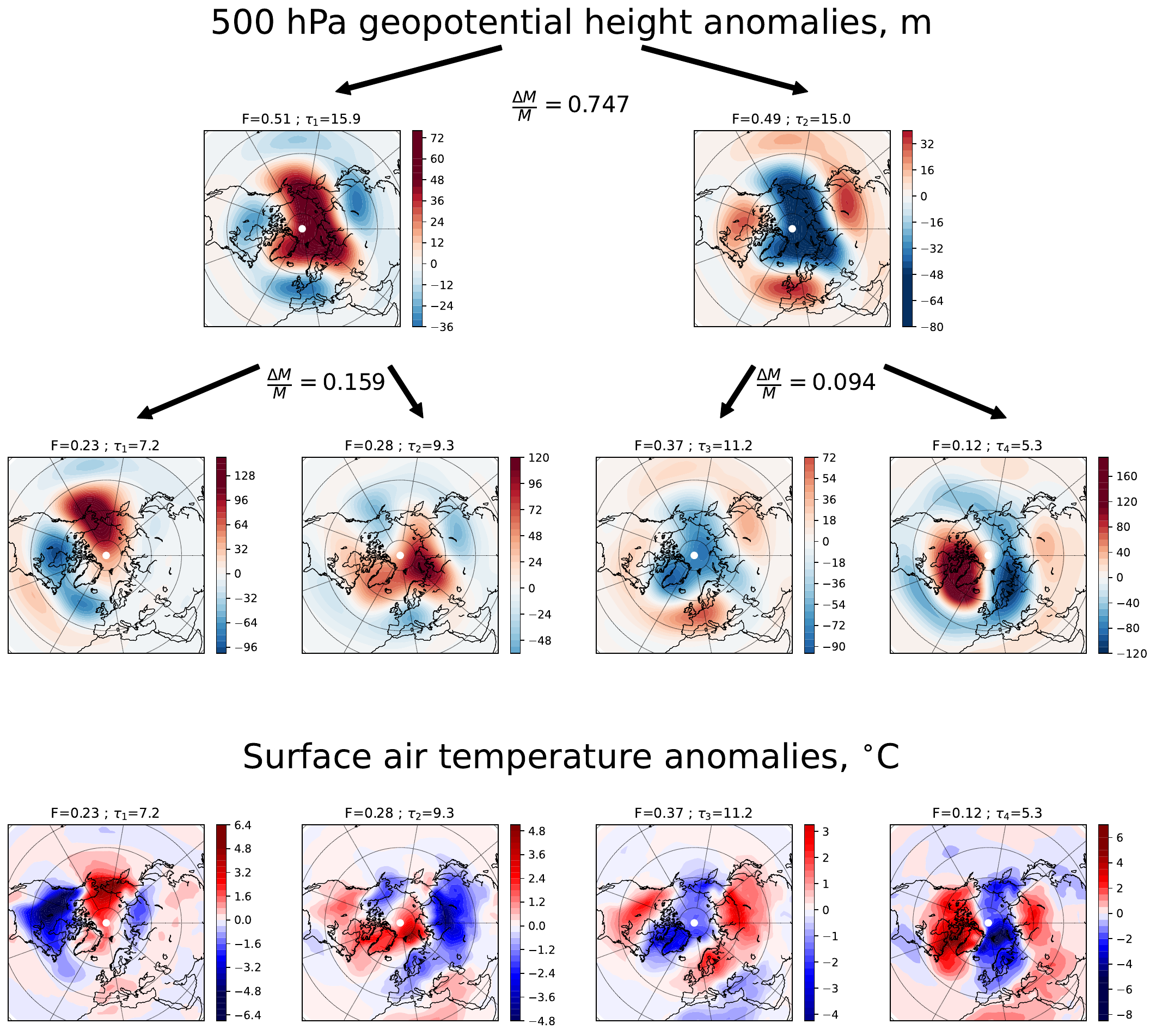}
\caption{Composite patterns of regimes obtained with $L=2$ days (see caption to  Fig.~3 in the main article) along with the corresponding surface temperature composite.}
\label{rean_patt2}
\end{figure*}

\begin{figure*}[ht]
\centering
\includegraphics[width=1.\textwidth]{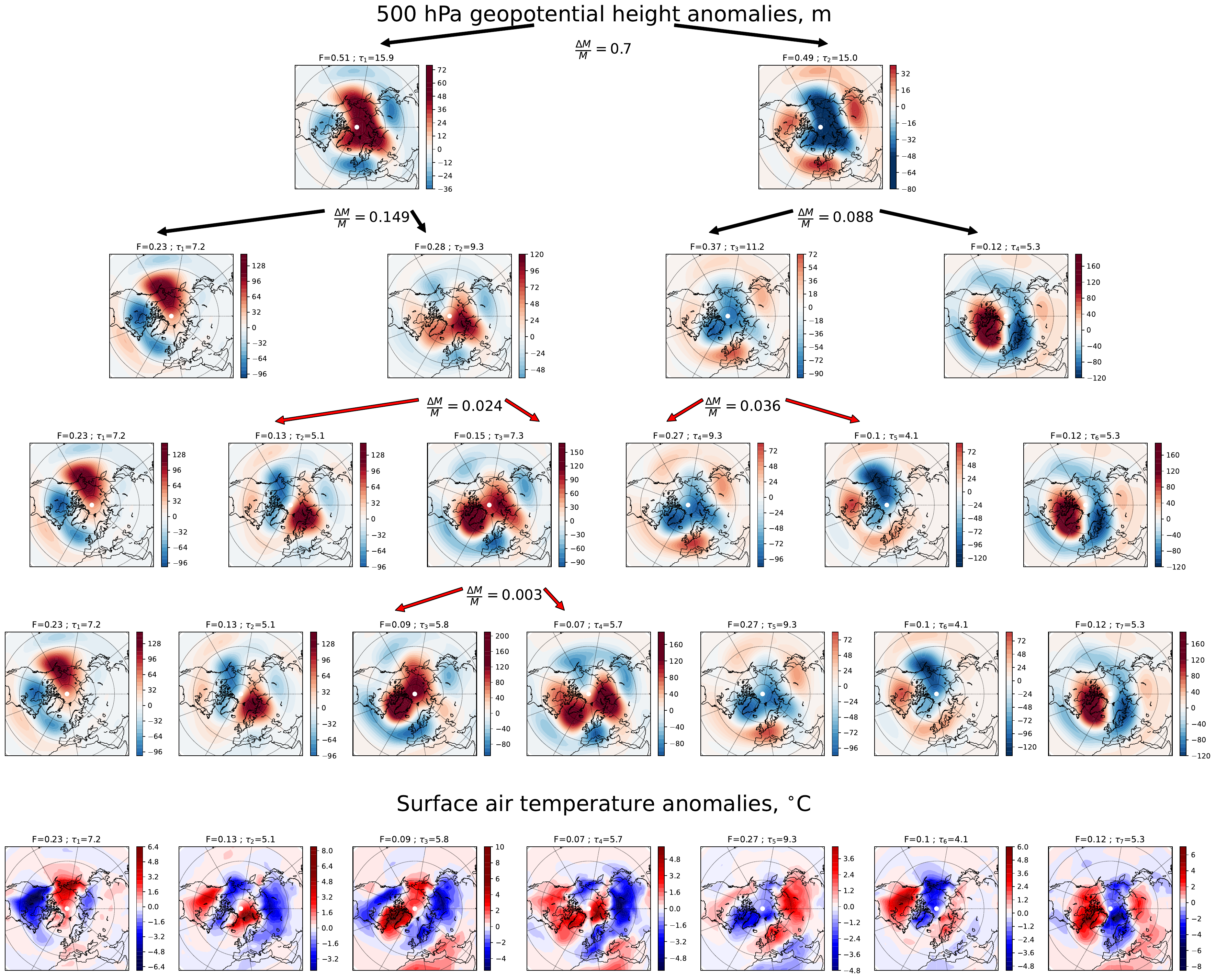}
\caption{Same as in Fig.~\ref{rean_patt2}, but complemented with splits that have not passed the significance test (red arrows).}
\label{comp_rean2}
\end{figure*}

\begin{figure*}[ht]
\centering
\includegraphics[width=1.\textwidth]{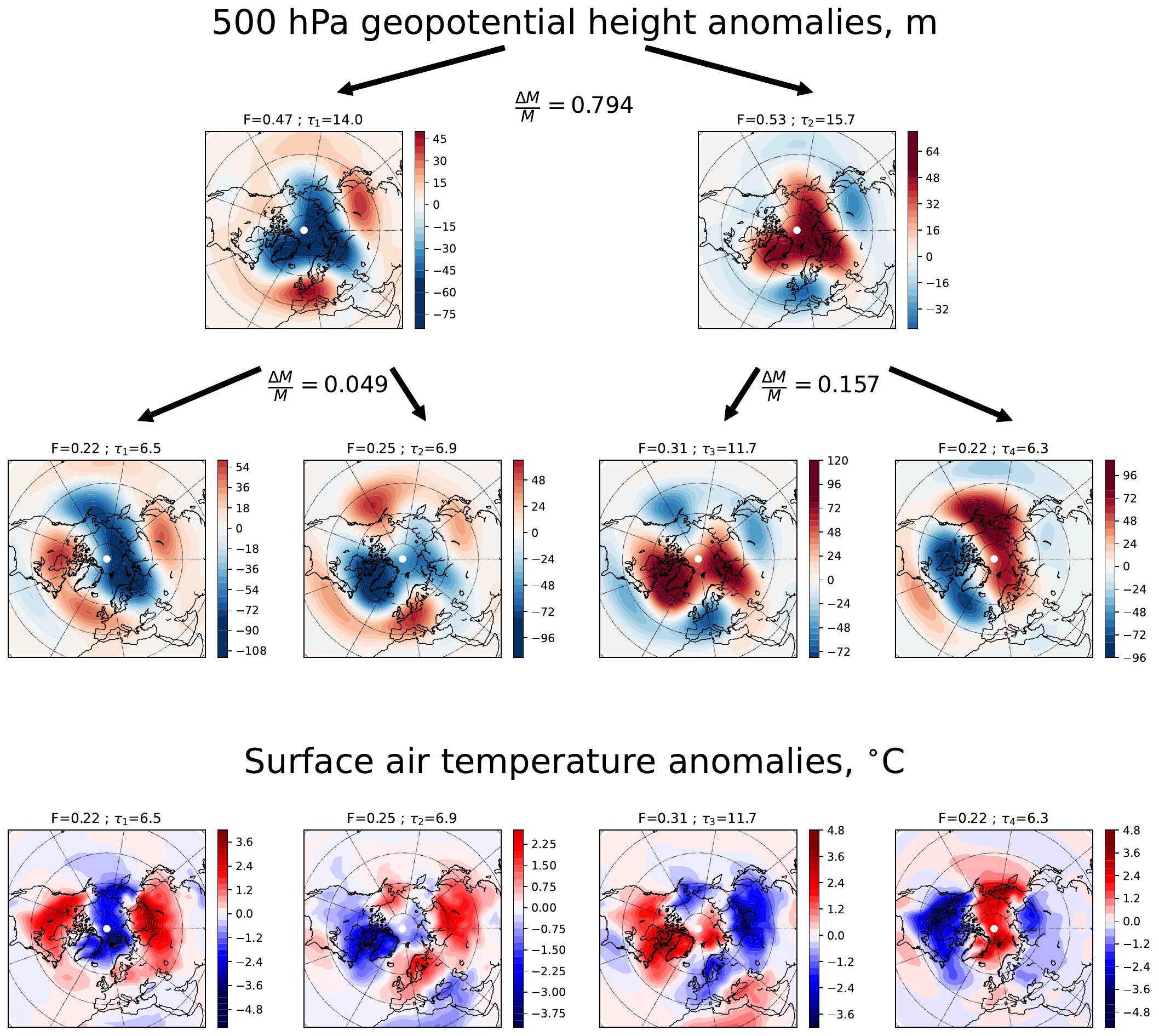}
\caption{Same as Fig. ~\ref{rean_patt2}, but with $L=4$.
 }
\label{rean_patt4}
\end{figure*}

\begin{figure*}[ht]
\centering
\includegraphics[width=1.\textwidth]{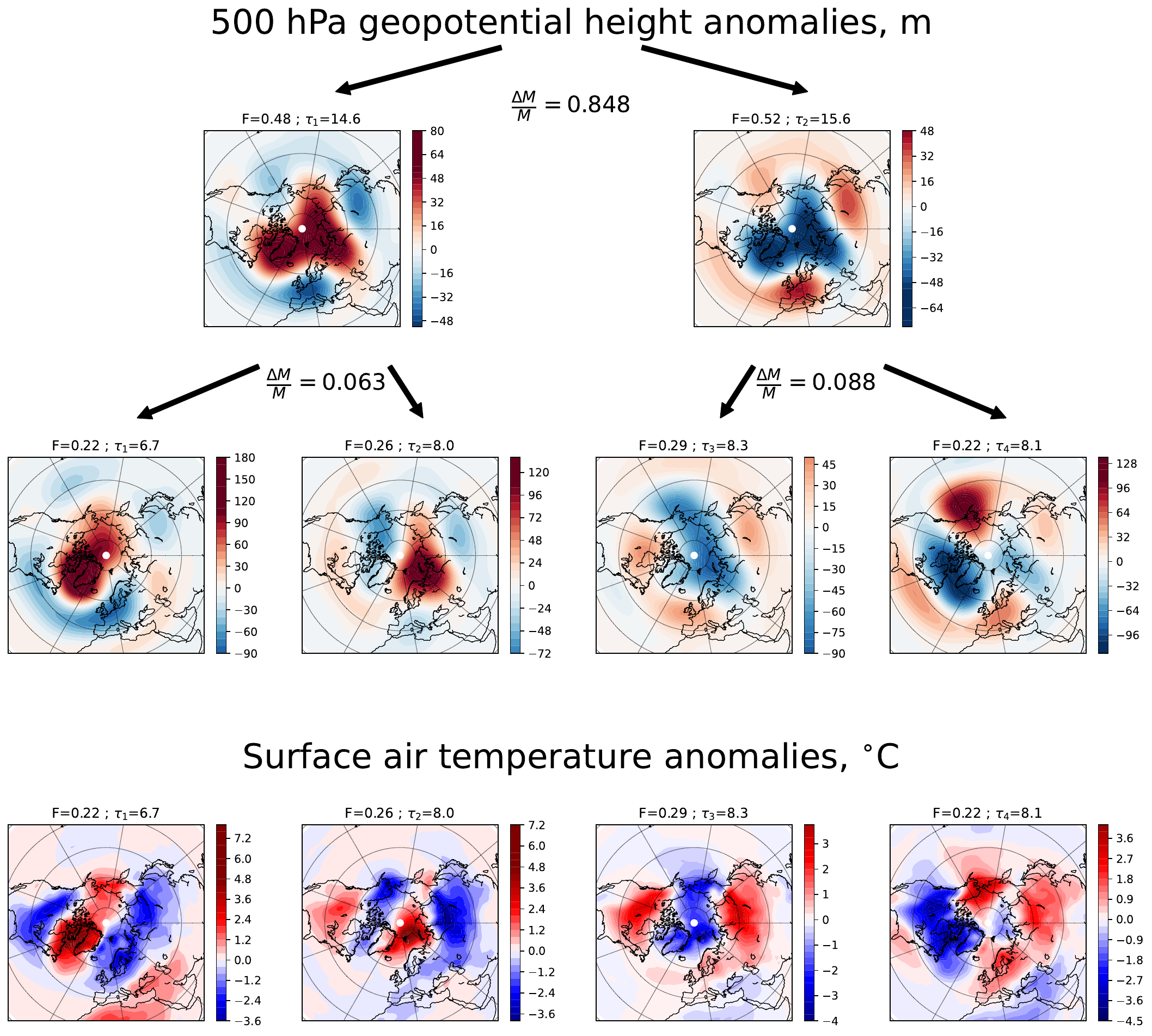}
\caption{Same as Fig. ~\ref{rean_patt2}, but with $L=6$.
 }
\label{rean_patt6}
\end{figure*}

\begin{figure*}[ht]
\centering
\includegraphics[width=1.\textwidth]{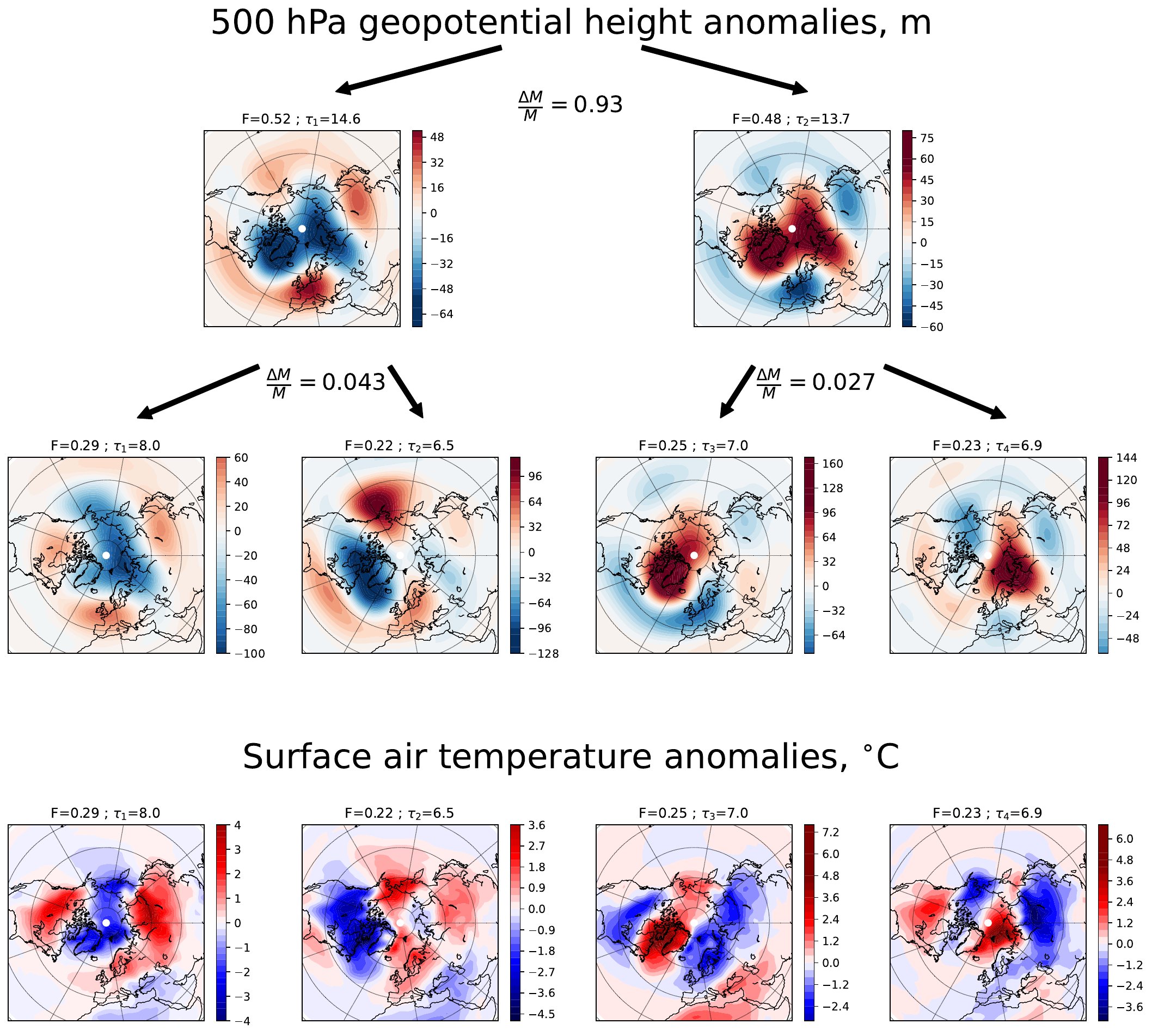}
\caption{Same as Fig. ~\ref{rean_patt2}, but with $L=8$.
 }
\label{rean_patt8}
\end{figure*}

\begin{figure*}[ht]
\centering
\includegraphics[width=1.\textwidth]{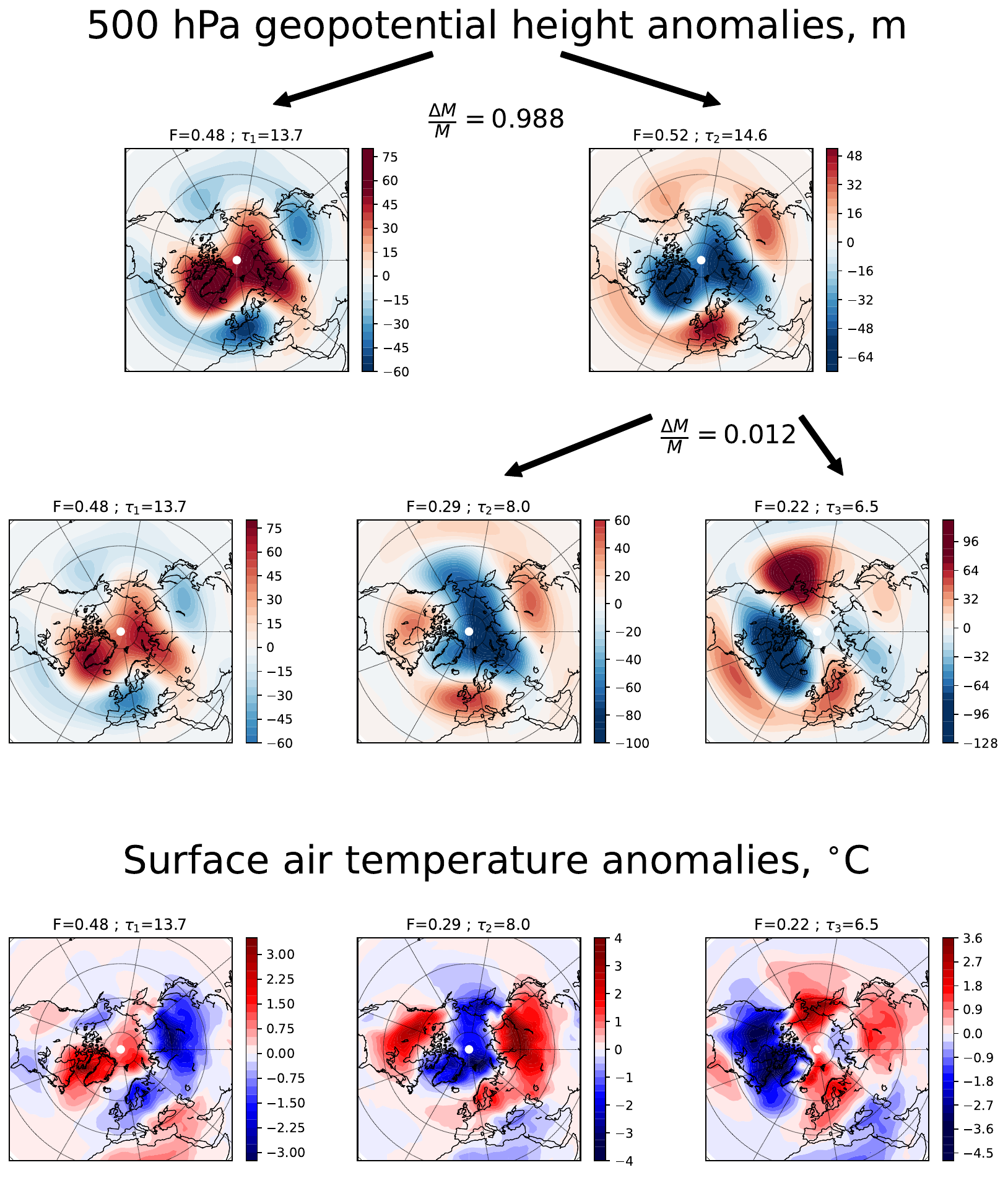}
\caption{Same as Fig. ~\ref{rean_patt2}, but with $L=10$.
 }
\label{rean_patt10}
\end{figure*}

\begin{figure*}[ht]
\centering
\includegraphics[width=1.\textwidth]{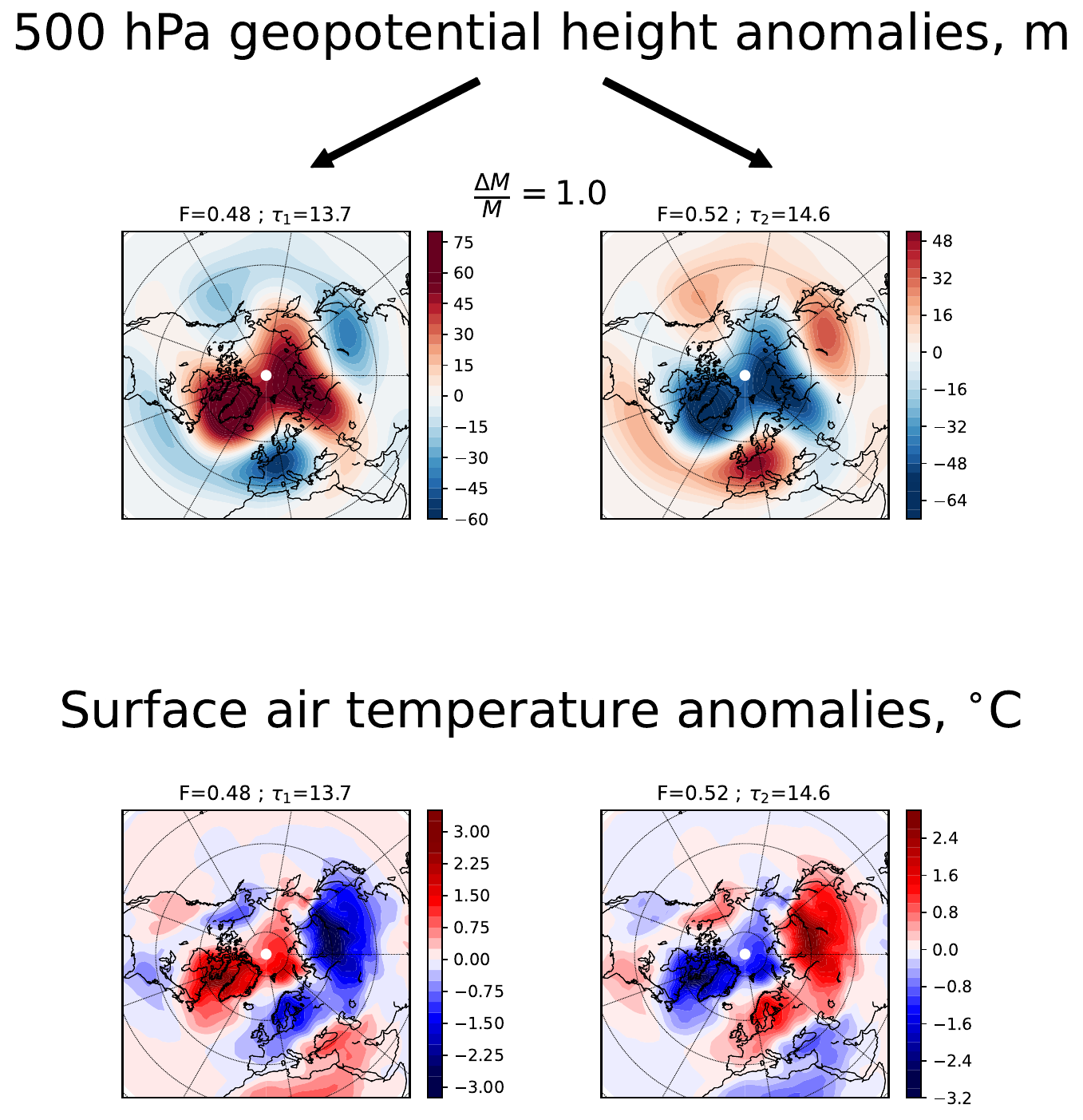}
\caption{Same as Fig. ~\ref{rean_patt2}, but with $L=11$.
 }
\label{rean_patt11}
\end{figure*}

%\bibliography{sm-scibib}

%\addtocounter{\@listctr}{8}
%\bibliography{sm-scibib}

%\bibliographystyle{Science}